\documentclass[11pt]{article}
\usepackage[dvips]{graphicx}
\usepackage{float}
\usepackage{epsfig}
\usepackage{ulem}
\usepackage{latexsym,amsmath,amsfonts,amssymb}
\usepackage[latin1]{inputenc}
\usepackage{rotating}
\usepackage[american]{babel}
\usepackage[dvips]{graphicx}
\usepackage{bbm}
\usepackage{color}
\usepackage{slashed}
\usepackage[unicode]{hyperref}
\usepackage{lscape}
\def\im{Invent. Math.}

\def\a{\alpha}
\def\b{\beta}
\def\c{\gamma}
\def\d{\delta}
\def\f{\phi}               
\def\vf{\varphi}  
\def\tvf{\tilde{\varphi}}
\def\vp{\varphi}
\def\g{\gamma}
\def\h{\eta}
\def\j{\psi}
\def\k{\kappa}                    
\def\l{\lambda}
\def\m{\mu}
\def\n{\nu}
\def\o{\omega}  \def\w{\omega}

\def\q{\theta}  \def\th{\theta}                  
\def\r{\rho}                                     
\def\s{\sigma}                                   
\def\t{\tau}
\def\u{\upsilon}
\def\x{\xi}
\def\z{\zeta}
\def\pt{\tilde{\varphi}}
\def\tt{\tilde{\theta}}
\def\lab{\label}
\def\6{\partial}
\def\wg{\wedge}
\def\bpsi{\bar{\psi}}
\def\bt{\bar{\theta}}
\def\bvf{\bar{\varphi}}

\DeclareMathOperator{\tr}{tr}

\newcommand{\be}{\begin{equation}}
\newcommand{\ee}{\end{equation}}
\newcommand{\beq}{\begin{equation}}
\newcommand{\eeq}{\end{equation}}
\newcommand{\bea}{\begin{eqnarray}}
\newcommand{\eea}{\end{eqnarray}}

\newcommand{\ba}{\begin{eqnarray}}
\newcommand{\ea}{\end{eqnarray}}

\newcommand{\beqs}{\begin{eqnarray}}
\newcommand{\eeqs}{\end{eqnarray}}
\newcommand{\bal}{\begin{aligned}}
\newcommand{\eal}{\end{aligned}}

\begin{document}
\baselineskip=15.5pt
\pagestyle{plain}
\setcounter{page}{1}

\def\del{{\partial}}
\def\vev#1{\left\langle #1 \right\rangle}
\def\cn{{\cal N}}
\def\co{{\cal O}}


\def\IC{{\mathbb C}}
\def\IR{{\mathbb R}}
\def\IZ{{\mathbb Z}}
\def\RP{{\bf RP}}
\def\CP{{\bf CP}}
\def\Poincare{{Poincar\'e }}
\def\tr{{\rm tr}}
\def\tp{{\tilde \Phi}}

\def\TL{\hfil$\displaystyle{##}$}
\def\TR{$\displaystyle{{}##}$\hfil}
\def\TC{\hfil$\displaystyle{##}$\hfil}
\def\TT{\hbox{##}}
\def\HLINE{\noalign{\vskip1\jot}\hline\noalign{\vskip1\jot}}
\def\seqalign#1#2{\vcenter{\openup1\jot
   \halign{\strut #1\cr #2 \cr}}}
\def\lbldef#1#2{\expandafter\gdef\csname #1\endcsname {#2}}
\def\eqn#1#2{\lbldef{#1}{(\ref{#1})}%
\begin{equation} #2 \label{#1} \end{equation}}
\def\eqalign#1{\vcenter{\openup1\jot
     \halign{\strut\span\TL & \span\TR\cr #1 \cr
    }}}

\def\eno#1{(\ref{#1})}
\def\href#1#2{#2}
\def\half{\frac{1}{2}}



\def\ads{{\it AdS}}
\def\adsp{{\it AdS}$_{p+2}$}
\def\cft{{\it CFT}}

\newcommand{\ber}{\begin{eqnarray}}
\newcommand{\eer}{\end{eqnarray}}

\newcommand{\beqar}{\begin{eqnarray}}
\newcommand{\cN}{{\cal N}}
\newcommand{\cO}{{\cal O}}
\newcommand{\cA}{{\cal A}}
\newcommand{\cT}{{\cal T}}
\newcommand{\cF}{{\cal F}}
\newcommand{\cC}{{\cal C}}
\newcommand{\cR}{{\cal R}}
\newcommand{\cW}{{\cal W}}
\newcommand{\eeqar}{\end{eqnarray}}
\newcommand{\tht}{\thteta}
\newcommand{\lm}{\lambda}\newcommand{\Lm}{\Lambda}


\newcommand{\nonu}{\nonumber}
\newcommand{\oh}{\displaystyle{\frac{1}{2}}}
\newcommand{\dsl}
   {\kern.06em\hbox{\raise.15ex\hbox{$/$}\kern-.56em\hbox{$\partial$}}}
\newcommand{\id}{i\!\!\not\!\partial}
\newcommand{\as}{\not\!\! A}
\newcommand{\ps}{\not\! p}
\newcommand{\ks}{\not\! k}
\newcommand{\D}{{\cal{D}}}
\newcommand{\dv}{d^2x}
\newcommand{\Z}{{\cal Z}}
\newcommand{\N}{{\cal N}}
\newcommand{\Dsl}{\not\!\! D}
\newcommand{\Bsl}{\not\!\! B}
\newcommand{\Psl}{\not\!\! P}

\newcommand{\eeqarr}{\end{eqnarray}}
\newcommand{\ZZ}{{\rm \kern 0.275em Z \kern -0.92em Z}\;}


\def\del{{\delta^{\hbox{\sevenrm B}}}} \def\ex{{\hbox{\rm e}}}
\def\azb{A_{\bar z}} \def\az{A_z} \def\bzb{B_{\bar z}} \def\bz{B_z}
\def\czb{C_{\bar z}} \def\cz{C_z} \def\dzb{D_{\bar z}} \def\dz{D_z}
\def\im{{\hbox{\rm Im}}} \def\mod{{\hbox{\rm mod}}} \def\tr{{\hbox{\rm Tr}}}
\def\ch{{\hbox{\rm ch}}} \def\imp{{\hbox{\sevenrm Im}}}
\def\trp{{\hbox{\sevenrm Tr}}} \def\vol{{\hbox{\rm Vol}}}
\def\rl{\Lambda_{\hbox{\sevenrm R}}} \def\wl{\Lambda_{\hbox{\sevenrm W}}}
\def\fc{{\cal F}_{k+\cox}} \def\vev{vacuum expectation value}
\def\nodiv{\mid{\hbox{\hskip-7.8pt/}}}
\def\ie{{\em i.e.}}
\def\ie{\hbox{\it i.e.}}

\def\CC{{\mathchoice
{\rm C\mkern-8mu\vrule height1.45ex depth-.05ex
width.05em\mkern9mu\kern-.05em}
{\rm C\mkern-8mu\vrule height1.45ex depth-.05ex
width.05em\mkern9mu\kern-.05em}
{\rm C\mkern-8mu\vrule height1ex depth-.07ex
width.035em\mkern9mu\kern-.035em}
{\rm C\mkern-8mu\vrule height.65ex depth-.1ex
width.025em\mkern8mu\kern-.025em}}}

\def\RR{{\rm I\kern-1.6pt {\rm R}}}
\def\NN{{\rm I\!N}}
\def\ZZ{{\rm Z}\kern-3.8pt {\rm Z} \kern2pt}
\def\IB{\relax{\rm I\kern-.18em B}}
\def\ID{\relax{\rm I\kern-.18em D}}
\def\II{\relax{\rm I\kern-.18em I}}
\def\IP{\relax{\rm I\kern-.18em P}}
\newcommand{\CS}{{\scriptstyle {\rm CS}}}
\newcommand{\CSs}{{\scriptscriptstyle {\rm CS}}}
\newcommand{\rc}{\nonumber\\}
\newcommand{\bear}{\begin{eqnarray}}
\newcommand{\eear}{\end{eqnarray}}

\newcommand{\LL}{{\cal L}}

\def\mani{{\cal M}}
\def\calo{{\cal O}}
\def\calb{{\cal B}}
\def\calw{{\cal W}}
\def\calz{{\cal Z}}
\def\cald{{\cal D}}
\def\calc{{\cal C}}

\def\to{\rightarrow}
\def\ele{{\hbox{\sevenrm L}}}
\def\ere{{\hbox{\sevenrm R}}}
\def\zb{{\bar z}}
\def\wb{{\bar w}}
\def\nodiv{\mid{\hbox{\hskip-7.8pt/}}}
\def\menos{\hbox{\hskip-2.9pt}}
\def\dr{\dot R_}
\def\drr{\dot r_}
\def\ds{\dot s_}
\def\da{\dot A_}
\def\dga{\dot \gamma_}
\def\ga{\gamma_}
\def\dal{\dot\alpha_}
\def\al{\alpha_}
\def\cl{{closed}}
\def\cls{{closing}}
\def\vev{vacuum expectation value}
\def\tr{{\rm Tr}}
\def\to{\rightarrow}
\def\too{\longrightarrow}


\def\a{\alpha}
\def\b{\beta}
\def\c{\gamma}
\def\d{\delta}
\def\e{\epsilon}           
\def\F{\Phi}
\def\f{\phi}               
\def\vf{\varphi}  \def\tvf{\tilde{\varphi}}
\def\vp{\varphi}
\def\g{\gamma}
\def\h{\eta}
\def\j{\psi}
\def\k{\kappa}                    
\def\l{\lambda}
\def\m{\mu}
\def\n{\nu}
\def\o{\omega}  \def\w{\omega}
\def\q{\theta}  \def\th{\theta}                  
\def\r{\rho}                                     
\def\s{\sigma}                                   
\def\t{\tau}
\def\u{\upsilon}
\def\x{\xi}
\def\X{\Xi}
\def\z{\zeta}
\def\pt{\tilde{\varphi}}
\def\tt{\tilde{\theta}}
\def\lab{\label}
\def\6{\partial}
\def\wg{\wedge}
\def\atanh{{\rm arctanh}}
\def\bpsi{\bar{\psi}}
\def\bt{\bar{\theta}}
\def\bvf{\bar{\varphi}}

%



\newfont{\namefont}{cmr10}
\newfont{\addfont}{cmti7 scaled 1440}
\newfont{\boldmathfont}{cmbx10}
\newfont{\headfontb}{cmbx10 scaled 1728}





\newcommand{\re}{\,\mathbb{R}\mbox{e}\,}
\newcommand{\hyph}[1]{$#1$\nobreakdash-\hspace{0pt}}
\providecommand{\abs}[1]{\lvert#1\rvert}
\newcommand{\Nugual}[1]{$\mathcal{N}= #1 $}
\newcommand{\sub}[2]{#1_\text{#2}}
\newcommand{\partfrac}[2]{\frac{\partial #1}{\partial #2}}
\newcommand{\bsp}[1]{\begin{equation} \begin{split} #1 \end{split} \end{equation}}
\newcommand{\calF}{\mathcal{F}}
\newcommand{\calO}{\mathcal{O}}
\newcommand{\calM}{\mathcal{M}}
\newcommand{\calV}{\mathcal{V}}
\newcommand{\bbZ}{\mathbb{Z}}
\newcommand{\bbC}{\mathbb{C}}
\newcommand{\cK}{{\cal K}}

\newcommand{\Thq}{\Theta\left(\r-\r_q\right)}
\newcommand{\Dq}{\d\left(\r-\r_q\right)}
\newcommand{\kten}{\kappa^2_{\left(10\right)}}
\newcommand{\pbi}[1]{\imath^*\left(#1\right)}
\newcommand{\ho}{\hat{\omega}}
\newcommand{\tth}{\tilde{\th}}
\newcommand{\tf}{\tilde{\f}}
\newcommand{\tj}{\tilde{\j}}
\newcommand{\tw}{\tilde{\omega}}
\newcommand{\tz}{\tilde{z}}
\newcommand{\prj}[2]{(\partial_r{#1})(\partial_{\j}{#2})-(\partial_r{#2})(\partial_{\j}{#1})}
\def\atanh{{\rm arctanh}}
\def\sech{{\rm sech}}
\def\csch{{\rm csch}}
\allowdisplaybreaks[1]

\def\red{\textcolor[rgb]{0.98,0.00,0.00}}

\newcommand{\Dan}[1] {{\textcolor{blue}{#1}}}

\numberwithin{equation}{section}

\newcommand{\Tr}{\mbox{Tr}}    


%

\setcounter{footnote}{0}
\renewcommand{\theequation}{{\rm\thesection.\arabic{equation}}}

\begin{titlepage}

\vspace{0.1in}

\begin{center}
\Large \bf Field Theory Aspects of non-Abelian T-duality and $\mathcal{N}=2$ Linear Quivers
\end{center}
\vskip 0.2truein
\begin{center}
 
Yolanda Lozano$^{a,}$\footnote{ylozano@uniovi.es} and Carlos N\'u\~nez$^{b,}$\footnote{c.nunez@swansea.ac.uk}

\vspace{.2in}
{\it $a$:Department of Physics, University of Oviedo,
Avda. Calvo Sotelo 18, 33007 Oviedo, Spain.}
\vskip 3mm
{\it $b$: Department of Physics, Swansea University, Swansea SA2 8PP, United Kingdom.}


\vspace{0.2in}
\end{center}
\vspace{0.2in}

\centerline{\bf Abstract:}
In this paper we propose a linear quiver with gauge groups of increasing rank
as 
field theory dual to the $AdS_5$ background constructed by Sfetsos and Thompson through non-Abelian T-duality.
The  formalism to study 4d ${\cal N}=2$ SUSY CFTs developed by Gaiotto and Maldacena is essential for our proposal.  
We point out an interesting relation between (Hopf) Abelian
and non-Abelian T-dual backgrounds that allows to see both backgrounds as different limits of a solution constructed by Maldacena and N\'u\~nez. This suggests different completions of the long quiver describing the CFT dual to the non-Abelian T-dual background that match different observables.

\smallskip
\end{titlepage}
\setcounter{footnote}{0}


\setcounter{footnote}{0}
\renewcommand{\theequation}{{\rm\thesection.\arabic{equation}}}
 \section{Introduction}
 Even in cases when a weakly coupled Lagrangian is not available, it  is possible to make precise statements about Quantum Field Theories (QFTs).
 When the theory is in a conformal phase, the constraints imposed by the symmetries allow to calculate and obtain exact results.
In two dimensions for example, the so called minimal models allow various analytic computations without any reference to a Lagrangian.
Four-dimensional Conformal Field Theories (4d CFTs) are not as constrained as those in two dimensions, but it is also possible to have an analytic understanding of the dynamics
 without a weakly coupled description, as long as there is some amount of supersymmetry. 
 A second way to deal with the problem of calculating observables for a 
 CFT at strong coupling (without the help of a Lagrangian) is to use a weakly 
 coupled string theory on an Anti de-Sitter space $AdS$ \cite{Maldacena:1997re}.  Different dualities -- like $SL(2,Z)$, T-duality-- characteristic of string theory, have become in this way 
common tools in the field theoretical understanding of CFTs.
 

In this paper we will study the interplay between T-duality and the AdS/CFT correspondence.   Abelian T-duality in $AdS$ string theory backgrounds is known to provide a mapping between different realisations of the same dual CFT,  but a clear understanding of this for its non-Abelian 
counterpart \cite{delaOssa:1992vci}, \cite{Giveon:1993ai}, 
\cite{Alvarez:1993qi} is still missing. Since it is not proven that non-Abelian  T-duality is a full string theory symmetry \cite{Alvarez:1993qi}, this opens  the exciting  possibility to construct new CFTs dual to $AdS$ string theory backgrounds, by the action of non-Abelian T-duality.

In the past four years, starting with the seminal paper \cite{Sfetsos:2010uq}, there has been quite some interest on the applications of non-Abelian T-duality to backgrounds
with a well understood QFT dual. Roughly, the procedure adopted was to consider a background with a well-known dual pair, where a non-Abelian T-duality  on a--typically $SU(2)$-- isometry was performed, generating
a new solution to the supergravity equations of motion. The calculation of  various field theoretical quantities with the new background, allowed for a partial exploration of the QFT associated to it.
This was complemented with an analysis of SUSY preservation and G-structures, for the initial and final solutions.
 This procedure was adopted in various works \cite{varios1}-\cite{Macpherson:2015tka}. 
 A more precise description of the QFT in terms of a quiver and a super-potential, was however mostly missing from these proposals. One of the technical points preventing a clear QFT interpretation of non-Abelian T-duality is the infinite range of the dual coordinates  \cite{Alvarez:1993qi}.
 
 In this paper we come back to the simple example of $AdS_5\times S^5$ to deal with this problem.
 The outcome of our study will be a proposal for a 4d ${\cal N}=2$ SUSY CFT dual to the
 background obtained by applying non-Abelian T-duality. 
 Exploiting the fact that this geometry fits in 
  the (Type IIA) classification 
 given by Lin, Lunin, Gaiotto and Maldacena in the papers \cite{Lin:2004nb}, \cite{Gaiotto:2009gz} we will make important use of the formalism developed there to propose a field theoretic description. 
 Reversing the logic, we will be able to identify the explicit Gaiotto-Maldacena geometry dual to a 4d $\mathcal{N}=2$ superconformal linear quiver with gauge groups of increasing rank in terms of the non-Abelian T-dual of $AdS_5\times S^5$ constructed in  \cite{Sfetsos:2010uq}.
 We will  explore as well different issues of the (Hopf) Abelian T-dual of $AdS_5\times S^5$. We will show that it fits in the previous classification of $\mathcal{N}=2$ geometries, thus providing a realisation of $\mathbb{Z}_n$ orbifolds of $AdS_5\times S^5$ (including the trivial case $n=1$) as Gaiotto-Maldacena geometries, related to wrapped M5-branes. An interesting connection between Abelian and non-Abelian T-dualities, mirrored by certain observables in the associated dual CFTs will be proposed. This will be crucial in testing the dual CFTs.
Even if both Abelian and non-Abelian T-dual solutions obtained starting from $AdS_5\times S^5$ are singular,
 this does not prevent our analysis to go through---the observables we compute are not
 afflicted by the singularity. Our analysis can be repeated  in other smooth backgrounds such as the ones studied in \cite{varios1}-\cite{Macpherson:2015tka}.

 
 
 
 As will be explained in detail below, our results indicate that T-duality
(Abelian and non-Abelian),
produces (as a generating technique) backgrounds in the 
Gaiotto-Maldacena class.
These are dual to CFTs involving long linear quivers at strong coupling.
Hence, we could apply non-abelian T-duality to a given Type II solution
obtained as a flow from a  Gaiotto-Maldacena background, to generate a new
one. This must be thought of as a dual description to a flow from a Gaiotto
CFT. This idea is also applicable to solutions with less SUSY and smaller
isometry group (in cases where the Gaiotto-Maldacena formalism does not
apply). This points to a way to understand QFT phenomena in cases where
the modern field theoretical techniques do not apply.

 The paper is organised as follows. In Section \ref{backgrounds}, we present the Abelian and non-Abelian T-dual backgrounds used in our investigations. We analyse quantised charges and  large gauge transformations. While most of the material in this section is present in the bibliography, 
 we will also point out an interesting new relation between the Abelian and non-Abelian T-dual solutions.
 In Section \ref{branerealisation}, we present the Hanany-Witten brane set-up
that encodes and  summarises the brane configurations associated to the charges.
In Section \ref{Gaiotto-Maldacena} we show  
that both backgrounds fit the LLM-classification  \cite{Lin:2004nb}, and write the potentials associated to their Gaiotto-Maldacena descriptions
 \cite{Gaiotto:2009gz}. Making use of the formalism in \cite{Gaiotto:2009gz} we then propose a quiver that describes the field theory dual of the non-Abelian T-dual background. 
 We comment on a field-theory inspired way of supplementing this background to have a finite range for the T-dual coordinate $r$, that otherwise would be unbounded.
Then, in Section \ref{QFT}, we discuss field theory aspects that can be read from our backgrounds. 
We focus the attention on the central charge, entanglement entropy and 't Hooft coupling of
our CFTs. The quivers that we propose are shown to  precisely match the values of these observables. Finally, in Section \ref{secciondeconstruction}, we relate our 4d quivers with those used in deconstructing six dimensional CFTs \cite{ArkaniHamed:2001ca},
\cite{ArkaniHamed:2001ie}. Conclusions and possible lines to develop in the future are given in Section \ref{concl}.

\section{Geometry} \label{backgrounds}
In this section, we lay out the two Type IIA backgrounds that motivate this  investigation.
We start by presenting the well-known IIB $AdS_5\times S^5$ background, to fix notation and conventions. 
The vielbein, metric and Ramond five-form read,
\bea
& & e^{x_i}= \frac{2 R}{L}dx_i,\;\; e^R=\frac{2L}{R} dR,\;\;e^{\alpha}=2L d\alpha,\;\; e^\beta= 2L \sin\alpha d\beta, \;\; e^i= L \cos\alpha \omega_i,\nonumber\\
& & \sqrt{2}\omega_1= \cos\psi d\theta +\sin\psi\sin\theta d\varphi, \;\sqrt{2}\omega_2=-\sin\psi d\theta +\cos\psi\sin\theta d\varphi,\;\sqrt{2}\omega_3=d\psi+\cos\theta d\varphi.\nonumber\\
& & ds^2=\frac{ 4 R^2}{L^2}dx_{1,3}^2 +\frac{ 4 L^2}{R^2}dR^2 + L^2 \Big[4 d\alpha^2
+4 \sin^2\alpha d\beta^2 + 2 \cos^2\alpha (\omega_1^2+\omega_2^2+\omega_3^2) \Big],\nonumber\\
& &  F_5=\frac{2}{g_s L^4}(e^{tx_1x_2x_3R} + e^{\alpha\beta123})=\frac{64}{g_s L}R^3 (1+*_{10}) dR\wedge dt\wedge dx_1\wedge dx_2\wedge dx_3.\label{ads5xs5}
\eea
The ranges of the angular coordinates are $[0,\frac{\pi}{2}]$ for $\alpha$; both $\beta$ and $\varphi$  vary in $[0,2\pi]$, $\theta$ ranges  in $[0,\pi]$ and $\psi$ in $ [0,4\pi]$.
We define the generic polyform $\hat{F}$, the  quantised Page charges and the constants needed for their calculation as,
\bea
 \hat{F}= F e^{-B_2},\;\;
Q_{Page, Dp}=\frac{1}{2\kappa_{10}^2 T_{Dp}}\int\hat{F}_{8-p};\;\;\; 2\kappa_{10}^2 T_{Dp}=
(2\pi)^{7-p}g_s \alpha'^{\frac{(7-p)}{2}}.\label{values}
\eea
In the following, we set $g_s=1$. We impose the D3-brane charge to be an integer $N_{3}$, which implies a quantisation for the size of the space $L$,
\bea
Q_{D3}=
\frac{1}{2\kappa_{10}^2 T_{D3}}\int_{\Sigma_5} {\hat F}_5=N_{3}\to  \frac{L^4}{\alpha'^2}=\frac{\pi N_{3}}{4}.\label{quantisationd3}
\eea

\subsection{The Abelian T-dual}

The first type IIA background that will feature in our discussion is the Abelian T-dual, in the direction $\psi$, of the  $AdS_5\times S^5$ background  in eq.(\ref{ads5xs5}). 
The T-dual coordinate, that we still denote by  $\psi$, has now periodicity $\pi$ \footnote{In the $\sigma$-model derivation of Abelian T-duality the periodicity of the dual variable is fixed by the condition  \cite{Rocek:1991ps}
\begin{equation}
\label{dualvar}
\int d\theta \wedge d{\tilde \theta}=(2\pi)^2.
\end{equation}
This implies that a $\theta$ variable with periodicity $2\pi$ is mapped to a ${\tilde \theta}$ with the same periodicity. When the dualisation is performed on a Hopf-fibre direction of periodicity $4\pi$ the dual variable has periodicity $\pi$.}.
Renaming $(\theta,\varphi)\to (\chi,\xi)$ and following the rules in
\cite{Bergshoeff:1995as}, we find, 
\bea
& & {ds^2}=\frac{4R^2}{L^2}dx_{1,3}^2 +\frac{4 L^2}{R^2}dR^2 
+4 L^2 \Big[d\alpha^2
+\sin^2\alpha d\beta^2 \Big]+\frac{\alpha'^2 d\psi^2}{L^2\cos^2\alpha} \!+
L^2\cos^2\alpha (d\chi^2 +\sin^2\chi d\xi^2).\nonumber\\
& & {B_2}=\alpha' \psi \sin\chi d\chi \wedge d\xi,\;
e^{-2{\Phi}}=\frac{ L^2 \cos^2\alpha}{\alpha'}, \; F_4= \frac{8 L^4}{\sqrt{\alpha'}} \cos^3\alpha \sin\alpha \sin\chi d\alpha\wedge d\beta \wedge d\chi\wedge d\xi. 
\label{ads5xs5td}
\eea
Here we have chosen a particular 
gauge for $B_2$.
This gauge will prove to be very useful for our discussion below.
The dilaton and the $g_{\psi\psi}$ 
component of the metric show that the background is singular at $2\alpha=\pi$, where the original $S^3$ in eq. (\ref{ads5xs5}) shrinks to a point. The supersymmetry is reduced. Indeed, the background in eq.(\ref{ads5xs5td}) is known to be $\mathcal{N}=2$ supersymmetric  
\cite{Fayyazuddin:1999zu}, and dual to 
the $\mathcal{N}=2$ realisation of $\mathcal{N}=4$ SYM with a hypermultiplet in the adjoint \cite{Witten:1998qj,Witten:1997sc}.

Quantising the Page charge of D4 branes one obtains,
\beq
Q_{D4}=\frac{1}{2\kappa_{10}^2 T_{D4}}\int_{\Sigma_4} {\hat F}_4= N_{4}\to \frac{L^4}{\alpha'^2}= \frac{\pi N_{4}}{2}.\label{charged4}
\eeq
The factor of two compared to eq.(\ref{quantisationd3}) is due to the different periodicities of the original and T-dual variables. It is common to absorb it through a redefinition of Newton's constant. 
There is also NS-five brane charge, obtained via the integration of $H_3=dB_2$ on the three manifold spanned by $S^2(\chi,\xi)$ and  the $\psi$-direction,
\bea
Q_{NS5}=\frac{1}{2\kappa_{10}^2 T_{NS5}}\int_{\Sigma_3}H_3= \frac{1}{4\pi^2\alpha'} \int_0^{2\pi} d\xi \int_0^\pi d\chi \sin\chi \int_0^{\pi} d\psi= 1.
\label{chargens5t}
\eea
Here we have used that $(2\pi)^5 g_s^2\alpha'^3 T_{NS5}=1$. Also, notice that the three manifold $\Sigma_3=[\psi,\chi,\xi]$ does not shrink at $\psi=0$ or $\psi=\pi$. 
If, on the other hand,  we allow the $\psi$-coordinate to vary in $[0,n\pi]$, 
going 
$n$-times over the circle of length 
$\pi$, we find that $N_{5}=n$, 
which tells us that one unit of NS5-brane charge is created in 
each of these turns. We can link this to the existence of large gauge transformations, as we will discuss in the next subsection. The resulting theory is the Abelian T-dual of $AdS_5\times S^5/\mathbb{Z}_n$, which has been shown to arise as the near horizon limit of a semi-localised M5, M5' system \cite{Fayyazuddin:1999zu,Alishahiha:1999ds} (with $N_{4}$ M5 and $n$ M5', in our notation) followed by reduction to IIA.

\subsection{The non-Abelian T-dual}

The second type IIA background that we will discuss 
is the non-Abelian T-dual of eq.(\ref{ads5xs5}), 
first worked out in \cite{Sfetsos:2010uq}. In the notation and conventions of
 \cite{Macpherson:2014eza} it reads,
\bea
& & {ds^2}=\frac{4R^2}{L^2}dx_{1,3}^2 +\frac{4 L^2}{R^2}dR^2 
+ L^2 \Big[4d\alpha^2
+4\sin^2\alpha d\beta^2 \Big]+\frac{\alpha'^2}{L^2\cos^2\alpha} 
dr^2 +\nonumber\\
& & 
+\frac{\alpha'^2L^2\cos^2\alpha r^2}{\alpha'^2 r^2 
+ L^4\cos^4\alpha}(d\chi^2 +\sin^2\chi d\xi^2).\nonumber\\
& & {B_2}=\frac{\alpha'^3 r^3}{\alpha'^2 r^2 +L^4 \cos^4\alpha}
\sin\chi d\chi \wedge d\xi;\;\;\; 
e^{-2{\Phi}}=\frac{ L^2 \cos^2\alpha}{\alpha'^3}
(L^4 \cos^4\alpha +\alpha'^2 r^2).\nonumber\\
& & {F_2}=\frac{8L^4    }{\alpha'^{3/2}}
\sin\alpha \cos^3\alpha d\alpha \wedge d\beta,\;\;  {F}_4= {B}_2\wedge {F}_2.
\label{ads5xs5natd}
\eea
As in the previous  background both the 
dilaton and the $g_{rr}$ 
component of the metric are singular at the point where the 
original $S^3$ shrinks, $2\alpha=\pi$. We expect once more a reduction 
of supersymmetry. Indeed, this background preserves  
$\mathcal{N}=2$ SUSY,  as shown in  \cite{Sfetsos:2010uq}. 

The Page charges read,
 \bea
 \label{chargesnatd}
& & Q_{D6}=\frac{1}{2\kappa_{10}^2 T_{D6}}\int_{\alpha,\beta}F_2=N_{6}\to \frac{L^4}{\alpha'^2}=\frac{N_{6}}{2};\\ 
 & &  Q_{D4}=\frac{1}{2\kappa_{10}^2T_{D4}}\int \hat{F}_4=0.\nonumber
 \eea
As usual, after a non-Abelian T-duality transformation, the value of $\frac{L^2}{\alpha'}$ in  
eq.(\ref{chargesnatd}) is incompatible with that in eq. (\ref{charged4}). 
The radius of $AdS_5$ must  then 
take a different value 
\footnote{In this example we could absorb the 
different factor through a redefinition of Newton's constant, 
as in the Abelian case. But 
this is not possible in general when there are more charges 
involved, as it is the case for instance in the $AdS_6$ and $AdS_4$ backgrounds discussed 
in \cite{Lozano:2013oma,Lozano:2014ata}.}. 
According to eq.(\ref{chargesnatd}), D3-branes should be replaced 
by D6-branes after the non-Abelian T-duality. As we discuss below, extending
 \cite{Macpherson:2015tka} (see also  \cite{Lozano:2013oma,Lozano:2014ata}), there is however more to this interpretation.  

As before, there is also NS5-brane charge associated to 
$H_3=dB_2$. In order to properly define this 
charge we need to know the range of the $r$-coordinate. 
However, in the absence of global information coming from 
the sigma model derivation (there is no analogue to the condition in eq.
(\ref{dualvar}) in the non-Abelian case), the coordinate
$r$ ranges in $\mathbb{R}^+$. This poses an obvious problem to 
the interpretation of the number $N_{5}$. 

In the papers
\cite{Lozano:2013oma,Lozano:2014ata} an argument was proposed 
to determine the range of the
coordinate $r$ in backgrounds like the one of eq.(\ref{ads5xs5natd}).
It uses the boundedness of the action for a 
fundamental Euclidean string that wraps a non-trivial two cycle. It amounts to impose that the
quantity, 
\beq
b_0=\frac{1}{4\pi^2\alpha'}\oint_{\Sigma_2} B_2 ,
\label{b0zz}
\eeq
is bounded, $b_0\in [0,1]$.
In the case of
the background in eq.(\ref{ads5xs5natd}), a non-trivial cycle is
\cite{Lozano:2013oma,Macpherson:2014eza},
\beq
\Sigma_2=[\chi,\xi],\;\;\; \alpha=\frac{\pi}{2},\label{2-manifold}
\eeq
since the geometry spanned by $\alpha$ and $S^2(\chi,\xi)$ close to $2\alpha=\pi$ is conformal to a singular cone with $S^2(\chi,\xi)$ boundary. 
Imposing that the integral in eq.(\ref{b0zz}) is quantised, $b_0=n$, for some integer $n$,  implies the existence of special values $r_n= n\pi$  for the $r$-coordinate, such that when 
$r$ varies in intervals $[r_n, r_{n+1}]$, 
a piece-wise continuous large gauge transformation on the $B_2$ potential 
\beq
B_2\to B_2-n\pi \alpha' \sin\chi d\chi \wedge d\xi,\label{xxxb}
\eeq
should be performed, bringing $b_0$ back to be valued in the $[0,1]$ interval.
This  large gauge transformation of the $B_2$-field, has an important effect on the Page charges, that change as,
\beq
\Delta Q_{D6}=0,\;\;\;\; \Delta Q_{D4}= - n N_{6},\label{xxxpage}
\eeq
implying  that D4 brane charge is created, $N_{4}=n N_{6}$, in absolute value, when we pass through $r_n=n\pi$ points. Hence, the charge of D4 branes is not
globally defined, but depends on the interval $[r_{n}, r_{n+1}]$ where we measure it.
Coming back to the quantisation of NS-five brane charge, we can take the manifold $\Sigma_3=[r,\chi,\xi]$ to integrate $H_3=dB_2$---in this case, we will also let the $r$-coordinate vary in $[0, n\pi]$. We find,
\bea
& & Q_{NS5}=\frac{1}{2\kappa_{10}^2 T_{NS5}}\int_{\Sigma_3}H_3= \frac{1}{4\pi^2 \alpha'} \int_0^{2\pi} d\xi \int_0^\pi d\chi \sin\chi \int_0^{n\pi} dr= n=N_{5}.
\label{chargens5NATD}
\eea
So, every time we cross a $r_n$ point a unit of NS-five brane charge is added. 
We will use and give a gauge theoretic interpretation to
 these results for the quantised charges in the next section.

Note that the present analysis of large gauge transformations holds as well in the Abelian T-dual background of 
eq.(\ref{ads5xs5td}). Indeed, in the gauge taken there for 
the $B_2$ field, there is also a non-trivial 2-cycle 
$\Sigma_2=[\chi,\xi]$ at $2\alpha=\pi$ where large gauge 
transformations can be defined. Thus, if we go $n$-times over the circle 
of length $\pi$ on which the $\psi$ variable ranges, 
a continuous transformation of parameter $n$, 
as in eq. (\ref{xxxb}), must be performed every time we cross 
a $[n\pi,(n+1)\pi]$ interval. In each of these turns a unit 
of NS5-brane charge is  added, as we mentioned around eq. (\ref{chargens5t}). The difference with the non-Abelian 
case is that the D4-brane charge remains the same in each  
interval. This will be an important observation when we compare 
the quivers associated to the Abelian and non-Abelian T-dual backgrounds 
in Section \ref{Gaiotto-Maldacena}.

\subsection{Relation between the Abelian and non-Abelian T-duals}\label{seccionrelacion}
As observed in \cite{Macpherson:2015tka} 
in a more general context, the $r\rightarrow\infty$ 
limit of the NS-NS sector of the non-Abelian T-dual solution 
in eq.(\ref{ads5xs5natd}),  
reduces to that of the Abelian T-dual solution of eq.(\ref{ads5xs5td}), 
with the identification $r=\psi$. 
In this section we make more concrete this relation, 
which will be very inspiring in order to elucidate the field theory dual to the non-Abelian solution.

First, one should notice that in the $r\rightarrow\infty$ 
limit the dilaton fields 
of both solutions differ by a factor of $r^2$. 
Recalling that the dilaton is determined by a 1-loop effect in T-duality, 
this factor is there to account precisely for the 
different integration measures that enter in the definition of 
the partition functions of the Abelian and non-Abelian T-dual 
$\sigma$-models, namely the measures
$\sin{\chi}{\cal D}\psi {\cal D}\chi {\cal D}\xi$ and
$r^2 \sin{\chi} {\cal D}r {\cal D}\chi {\cal D}\xi$, respectively.
Secondly, the observation above 
that $r$ must be divided in intervals of length $\pi$ in order to properly account for large gauge transformations, 
allows us to identify $r$ and $\psi$ globally for $r\rightarrow\infty$ if we take $r$ in a  
$[n\pi,(n+1)\pi]$ interval and then send $n$ to infinity. 
We will see in Section \ref{QFT} that 
different quantities associated to the field theories dual to 
both  Abelian and non-Abelian backgrounds ( when computed with $r$ in the $[n\pi,(n+1)\pi]$ interval 
for the non-Abelian solution) will indeed agree in this limit.

Let us turn to the analysis of the RR sector. 
Clearly, the fluxes of the Abelian and non-Abelian T-dual solutions in 
eqs.(\ref{ads5xs5td}) and (\ref{ads5xs5natd}), 
are not the same in the $r\rightarrow\infty$ limit. 
There is however a neat relation between the associated quantised charges, 
that allows to conclude that both backgrounds are still physically 
equivalent in this limit. 

To be concrete, consider the non-Abelian T-dual background 
and look at the quantised charges defined 
in the $r\in [n\pi,(n+1)\pi]$ 
interval,
where the matching with the Abelian T-dual solution is expected to occur. 
In this interval, using eqs.(\ref{chargesnatd}) and (\ref{xxxpage}), we have
\begin{equation}
N_{6}= 2\frac{L^4}{ \alpha^{\prime 2}},\, \qquad N_{4}= n N_6.
\end{equation}
For the Abelian background we have, in turn
\begin{equation}
N_{4}=\frac{2}{\pi} \frac{L^4}{ \alpha^{\prime 2}}.
\end{equation}
Thus,
\begin{equation}
N_{4}^{NATD}=n\pi N_{4}^{ATD}\, ,
\label{relND4}
\end{equation}
and the factor of difference can be safely absorbed through a redefinition of Newton's constant.
We have checked that this same type of rescaling relates the charges of 
the non-Abelian and Abelian T-duals of other $AdS$ backgrounds where more quantized charges are 
present. 
The expectation is 
that this relation will hold more generally, 
and that we will be able to always absorb 
the $n\pi$ factor through a redefinition of Newton's constant \cite{LMN}.

We will now discuss aspects of the CFTs dual to each of our backgrounds. We will start by pointing out the relation with 
Hanany-Witten set-ups  \cite{Hanany:1996ie}.



\section{Brane realisation}\label{branerealisation}

\begin{figure}
\centering
\includegraphics[scale=0.5]{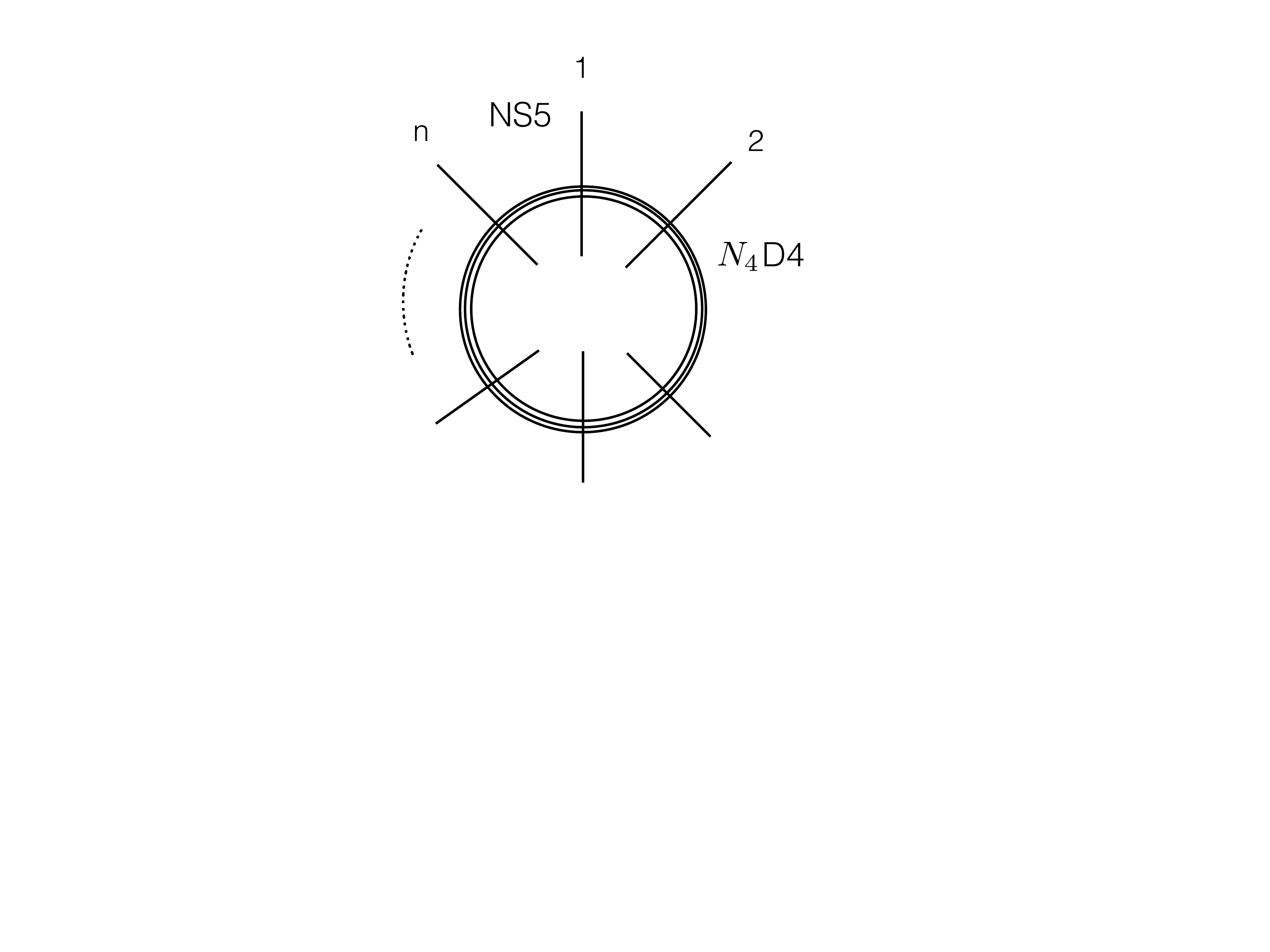}
\vspace{-7.5cm}
\caption{Brane set-up for the Abelian T-dual of $AdS_5\times S^5/\mathbb{Z}_n$. At each interval there are $N_{4}$ D4-branes.}
\label{braneab}
\end{figure}

In this section, we elaborate on a brane picture consistent with our previous findings for the quantised charges. Let us first recall that the brane set-up describing the Abelian T-dual of $AdS_5\times S^5$ is known in the literature. It consists on a set of $N_4$ D4-branes stretched between two NS5-branes that are identified \cite{Witten:1997sc}. The Abelian T-dual of $AdS_5\times S^5/\mathbb{Z}_n$ is in turn associated to a periodic array of NS5 and D4 branes \cite{Witten:1997sc,Alishahiha:1999ds}
as depicted in Figure \ref{braneab}.

\begin{figure}
\centering
\includegraphics[scale=0.5]{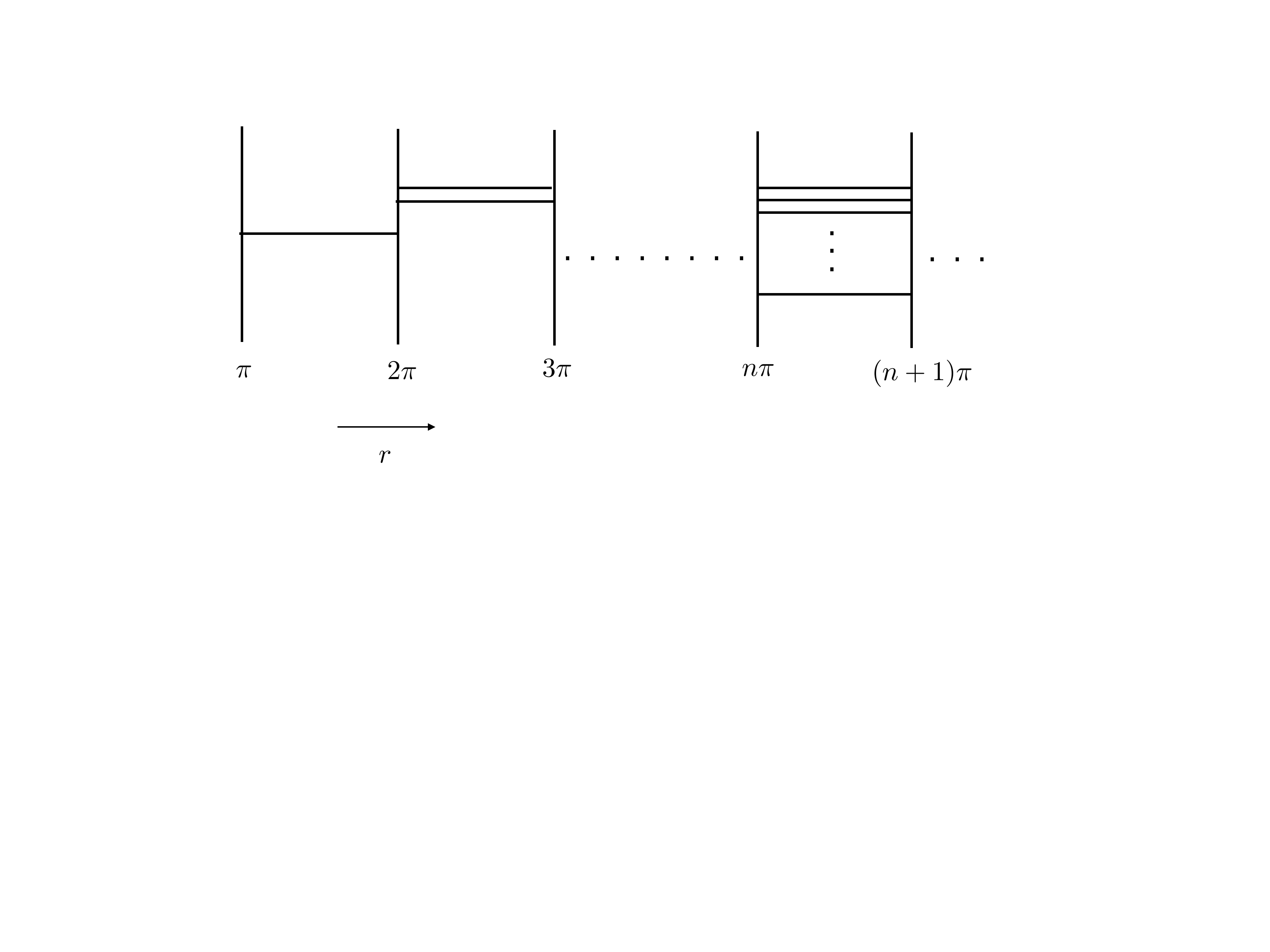}
\vspace{-7.5cm}
\caption{Brane set-up for the non-Abelian T-dual background. Parallel vertical branes are NS5-branes and horizontal branes are D4-branes suspended between them. At each $[n\pi,(n+1)\pi]$ interval there are $nN_6$ D4-branes.}
\label{branenonab}
\end{figure}

Let us now move to the more interesting non-Abelian case. It was discussed in \cite{Macpherson:2015tka} ---see eq.(3.16) of that paper-- that the non-Abelian T-dual of $AdS_5\times S^5$ should be related to a D4/NS5 brane set-up. Indeed, the analysis of the fluxes in eq.(\ref{ads5xs5natd}) and the quantised charges in eq.(\ref{chargesnatd}) suggests that we are dealing with D4 branes extended on $(\mathbb{R}^{1,3}, r)$ and NS-five branes extended along ($\mathbb{R}^{1,3},\alpha,\beta$). The NS5 branes are localised at positions
$r_n=n\pi$, as we learnt in eq.(\ref{chargens5NATD}). At each interval $[r_n, r_{n+1}]$ there are $nN_6$ D4-branes stretched in the $r$-direction. These branes can generate $N_6$ D6 branes extended along $(\mathbb{R}^{1,3},r,\chi,\xi)$ through Myers dielectric effect \cite{Myers:1999ps}. Here we will use D4-branes as colour branes. This will fit the Gaiotto-Maldacena description in Section \ref{Gaiotto-Maldacena}, and will also allow 
the matching with the Abelian result. In the $[0,\pi]$ interval the description should be, in turn, in terms of D6 branes, which could be thought of as a strong coupling effect, 
as we discuss in Section \ref{QFT}.  
The D4/NS5 brane set-up from $\pi$ onwards is summarised in Figure \ref{branenonab}.

The previous configuration is very resemblant of Witten's D4/NS5 brane set-up in  \cite{Witten:1997sc}, that we briefly summarise.  Witten presented a Type  IIA and M-theoretic  brane picture for 4d ${\cal N}=2$ field theories with 
a given number of gauge groups connected by bifundamental fields. The set-up contains NS-five branes extended
 in the directions $(\mathbb{R}^{1,3}, x_4,x_5)$ at different positions  $x_{6,n}$ in the $x_6$-direction, D4 branes extending in $(\mathbb{R}^{1,3}, x_6)$ in between the five branes and, possibly, D6 branes that cover the $(\mathbb{R}^{1,3},x_7,x_8,x_9)$ directions. A five dimensional gauge theory is realised on the four branes, but  having these finite extension in the $x_6$-direction, there is  a suitable low-energy limit (small energies compared to the inverse size of the D4 along $x_6$) in which this field theory is effectively four dimensional. The effective gauge coupling  behaves as $\frac{1}{g_4^2}\sim \frac{x_{6,n+1}- x_{6,n}}{g_s \sqrt{\alpha'}}$. In  \cite{Witten:1997sc}, it was
explained how the branes bend, making $x_6$ a function of  the coordinates $(x_4,x_5)$. The beta function is encoded in this 
functional dependence  of $x_6=f(x_4+ix_5)$, which is obtained by solving a Laplace equation.  The open strings connecting D4 branes stretched between different sets of NS5-branes  represent bifundamental matter. The strings connecting these D4's to either D6 branes or semi-infinite D4 branes (in the low energy limit above these have their excitations 'decoupled', acting like a global $SU(N_{6})$ group), represent the fundamental  matter. 
For $l_n$ D4-branes stretched between the $n^{th}$ and $(n+1)^{th}$ NS5-branes, with $n=1,\dots, N$, the gauge group is 
$\prod_{n=1}^{N} SU(l_n)$ ( the $U(1)'s$ are actually frozen, as discussed in \cite{Witten:1997sc}), and there are $(l_n,l_{n+1})$ hypermultiplets, that contribute to the $SU(l_n)$ $\beta$-function as $l_{n-1}+l_{n+1}$ flavors. Each $SU(l_n)$ gauge group has thus associated a coefficient for the $\beta$-function,
\begin{equation}
\label{beta}
b_{0,n}=-2l_n+l_{n+1}+l_{n-1}.
\end{equation}
This vanishes at each interval if $2l_n=l_{n+1}+l_{n-1}$.
Note that it is necessary to introduce semi-infinite D4 branes ending at the limiting NS5 branes, or D6-branes extended in $(\mathbb{R}^{1,3},x_7,x_8,x_9)$,  to account for the necessary fundamentals of the $SU(l_1)$ and $SU(l_N)$ gauge groups.

Many interesting developments came from this brane picture and its lift to M-theory. For example, in eleven dimensions, the theta-angle is realised as periodic translations in $x_{11}$, that is $\Theta\sim (x_{11,n+1}- x_{11,n})$.  Higgs branches can be studied and a correspondence between the Seiberg-Witten curve and the 'shape' of the
branes was developed.

If we compare the previous  set-up with our proposed brane configuration  for the non-Abelian T-dual of $AdS_5\times S^5$ (see Figure \ref{branenonab}), we see that both are clearly the same if we identify the $r$-coordinate of our non-Abelian configuration with the $x_6$-direction of Witten's brane set-up. The NS5 branes are localised at positions
$r_n=x_{6,n}=n\pi$ and we can also identify $(x_4,x_5)\sim (\alpha,\beta)$. The number of D4 branes at each $[r_n, r_{n+1}]$ interval is $l_n=n N_6$, and the beta function coefficient in eq.(\ref{beta}) clearly vanishes.

 In Section \ref{Gaiotto-Maldacena}, we will propose a conformal quiver that can be put in correspondence with this brane set-up. An intuitive way to describe our brane set-up and quiver is to start with an array of parallel  NS-five branes, with a D6 flavour brane on one of the ends of the array and move this D6 brane across the five branes towards the other end. Defining a linking number associated with the NS-five brane that has $r_{6}$ ($l_{6}$) D6 branes to its right (left) and $R_{4}$ ($L_{4}$)
D4 branes to its right (left) respectively,
\begin{equation}
L_{NS}=\frac{1}{2}(r_{6}-l_{6})+L_{4}-R_{4},\nonumber
\eeq
we find that $L_{NS}=-1/2$ for all NS5-branes. Thus,
 the Hanany-Witten brane creation effect  \cite{Hanany:1996ie}, associated with moving a flavour D6 across
the different NS-five branes, can be used to give  an explanation for the $n$ (created) D4-branes in a $n^{th}$ interval.

We will now make contact between our geometries in eqs.(\ref{ads5xs5td}), (\ref{ads5xs5natd}) and the formalism developed by Gaiotto and Maldacena in
 \cite{Gaiotto:2009gz}, to study the CFTs presented by Gaiotto in \cite{Gaiotto:2009we}. This formalism was developed to study ${\cal N}=2$ CFTs associated to the brane set-up discussed above.

\section{Abelian and non-Abelian T-duals as Gaiotto-Maldacena geometries}\label{Gaiotto-Maldacena}

In this section we show that both $\mathcal{N}=2$ 
backgrounds in eqs.(\ref{ads5xs5td}) and (\ref{ads5xs5natd}) 
are examples of Gaiotto-Maldacena backgrounds \cite{Gaiotto:2009gz}.
In the Abelian case this provides an explicit realisation of a $\mathbb{Z}_n$ orbifold of $\mathcal{N}=4$ SYM as a Gaiotto-Maldacena geometry in Type IIA. In the non-Abelian case it will give an important hint for constructing the quiver describing its dual CFT.

In general, these  $\mathcal{N}=2$ SUSY-preserving backgrounds with an $AdS_5$ factor, 
can be written in terms of a potential function $V(\sigma,\eta)$. 
In particular, denoting $V'=\partial_{\eta} V$ 
and $\dot{V}=\sigma\partial_{\sigma}V$, 
one can write the Type IIA 
generic Gaiotto-Maldacena solution as \cite{ReidEdwards:2010qs,Aharony:2012tz},
\bea
& & ds_{IIA,st}^2=\alpha'(\frac{2\dot{V} -\ddot {V}}{V''})^{1/2}
\Big[  4 AdS_5 +\mu^2\frac{2V'' \dot{V}}{\Delta} 
{d \Omega^{2}_2(\chi,\xi)}+\mu^2\frac{2V''}{\dot{V}}  
(d\sigma^2+d\eta^2)+ \mu^2\frac{4V'' \sigma^2}{2\dot{V}-\ddot{V}} 
d{\beta}^2 \Big], \nonumber\\
& & A_1=2\mu^4\sqrt{\alpha'}
\frac{2 \dot{V} \dot{V'}}{2\dot{V}-\ddot{V}}d{\beta},\;\;\;\; 
e^{4\Phi}= 4\frac{(2\dot{V}-\ddot{V})^3}{\mu^{4}V'' \dot{V}^2 \Delta^2}, 
\quad {\Delta = (2 \dot{V} - \ddot{V}) V'' + (\dot{V}')^2} \ ,  \nonumber \\
& & B_2=2\mu^2\alpha' (\frac{\dot{V} \dot{V'}}{\Delta} -\eta) 
d\Omega_2,\;\;\; {C}_3={-} 4\mu^4 \alpha'^{3/2}
\frac{\dot{V}^2 V''}{\Delta}d{\beta} \wedge d\Omega_2.
\label{metrica}
\eea
The radius of the space is $\mu^2\alpha'=L^2$. The  two-sphere $d \Omega^{2}_2(\chi,\xi)$ is parametrised by 
the angles $\chi$ and $\xi$ with corresponding volume 
form $d\Omega_{2}= \sin\chi d\chi \wedge d\xi$. The usual definition
$F_4= dC_3 + A_1\wedge H_3$ is also used.
The problem of writing  IIA/M-theory  solutions in this class, reduces
to  finding the function $V(\sigma,\eta)$ that solves
a Laplace equation with a given charge density $\lambda(\eta)$,
\beq
\partial_\sigma[\sigma \partial_\sigma V]+\sigma \partial^2_\eta V=0,\;\;\;\;\;\;\;\lambda(\eta)= \sigma\partial_\sigma V(\sigma,\eta)|_{\sigma=0}\; .\label{ecuagm}\nonumber
\eeq
Interestingly, the background and fluxes depend on $\dot{V}$,
$\dot{V}'$, $\dot{\dot{V}}$ and $V''=-\sigma^{-2}\ddot{V}$. 
Hence, given $\dot{V}$, we have all that is needed to write the 
Type IIA background. 
Like in any other problem described by a differential equation in partial derivatives, boundary conditions must be imposed.
Gaiotto and Maldacena found these conditions by enforcing a correct quantisation of charges and the smooth-shrinking of some
sub-manifolds. These conditions have been nicely summarised in \cite{ReidEdwards:2010qs},
\cite{Aharony:2012tz}, they are
\begin{itemize}
\item{$\dot{V}(\sigma=0,\eta)=\lambda(\eta)$ must vanish at $\eta=0$.}
\item{$\lambda(\eta)$ must be a piecewise linear continuous function, composed of segments of the form $\lambda=a_i\eta+q_i$, with $a_i$ an integer.}
\item{The change in the slope of two consecutive kinks must be a negative integer, $a_i-a_{i-1}<0$. A kink in which the gradient changes by $k$ units is associated with D6 branes or $A_{k}$ singularities in the M-theory lift.}
\item{The positions of the kinks must be at integer values in the $\eta$-axis}.
\item{Some solutions satisfy $\lambda(N_*)=0$. In this case, the $\eta$-coordinate is bounded in $[0,N_*]$. The associated electrostatic problem consists of a line of charge density $\lambda(\eta)$, bounded by two 'conducting plates' at the points $\eta=0, \eta=N_*$.}
\end{itemize}
From the previous electrostatic problem it is possible to read-off the quiver associated to the $\mathcal{N}=2$ dual CFT. As summarised in \cite{ReidEdwards:2010qs}, \cite{Aharony:2012tz},
\begin{itemize}
\item An $SU(n_i)$ gauge group is associated to each integer value of $\eta=\eta_i$, with the rank $n_i$ given by the value of the charge density at that point, $\lambda(\eta_i)=n_i$.
\item A kink in the line profile corresponds to extra $k_i$ fundamentals attached to the gauge group at the node $n_i$.
\end{itemize}

Finally, let us recall the Maldacena-N\'u\~nez background written in \cite{Maldacena:2000mw}. 
It was shown in \cite{ReidEdwards:2010qs} that this background can be considered as a fundamental building block from which many $\mathcal{N}=2$ Type IIA solutions can be constructed. We will show that this is the case for both our Abelian and non-Abelian T-dual solutions.
In the Gaiotto-Maldacena formalism, the Maldacena-N\'u\~nez background is described by a potential ${V}_{MN}(\sigma,\eta)$ whose derivative and associated charge density take the simple expression,
\bea
&&\sigma \partial_\sigma V_{MN}= \dot{V}_{MN}(\sigma,\eta)=\frac{1}{2}\Big[\sqrt{(N_c+\eta)^2 +\sigma^2} -\sqrt{(N_c-\eta)^2+\sigma^2}   \Big],\label{vdotmn}\\
&&\lambda_{MN}(\eta)=\frac{1}{2}\Big(|\eta+N_c| -|\eta-N_c|     \Big).\nonumber
\eea

With these pieces of the formalism in place, we will show that our Abelian and non-Abelian T-dual backgrounds fit in. We  shall also discuss the connection between $V_{MN}$
and the potentials describing the Abelian and non-Abelian T-dual backgrounds. To this we turn now.

\subsection{The case of the Abelian T-dual of $AdS_5\times S^5$}
For the Abelian T-dual background in eq.(\ref{ads5xs5td}),  
after redefining 
\beq
\psi=r=\frac{2L^2}{\alpha'}\eta,\;\; \sigma=\sin\alpha,
\label{redefinition}
\eeq 
the potential $V(\sigma,\eta)$ and charge density $\lambda(\eta)$ are found to be,
\beq
V_{ATD}=\log\sigma-\frac{\sigma^2}{2} +\eta^2;\;\;\;\; \lambda(\eta)=1.
\label{potentialdensityTD}
\eeq
Scaling the metric as in \cite{Gaiotto:2009gz} we find $\lambda(\eta)=N_{4}$, with $N_{4}$ the number of D4-branes that create the background.
It corresponds to the charge profile (a) in Figure \ref{profiles}.
Though this charge density does not satisfy the boundary condition $\lambda (\eta=0)=0$,  this case is still compatible
with the quivers depicted in Figure \ref{abelianquivers}, describing  the $\mathbb{Z}_n$ orbifold  of $\mathcal{N}=4$ SYM using $\mathcal{N}=2$ language \cite{Witten:1997sc}.  Indeed, in this case the $\eta$-direction is periodic, hence we do not need to impose the conditions mentioned above, in particular, we do not have to impose
 the condition $\lambda(0)=0$.

Note that the number of flavours and the number of colours satisfy $N_f=2N_4=2N_c$ for each node in the circular quiver depicted in Figure \ref{abelianquivers}, leading to a vanishing beta function in correspondence with the $AdS_5$ factor in the geometry.

\begin{figure}
\centering
\includegraphics[scale=0.45]{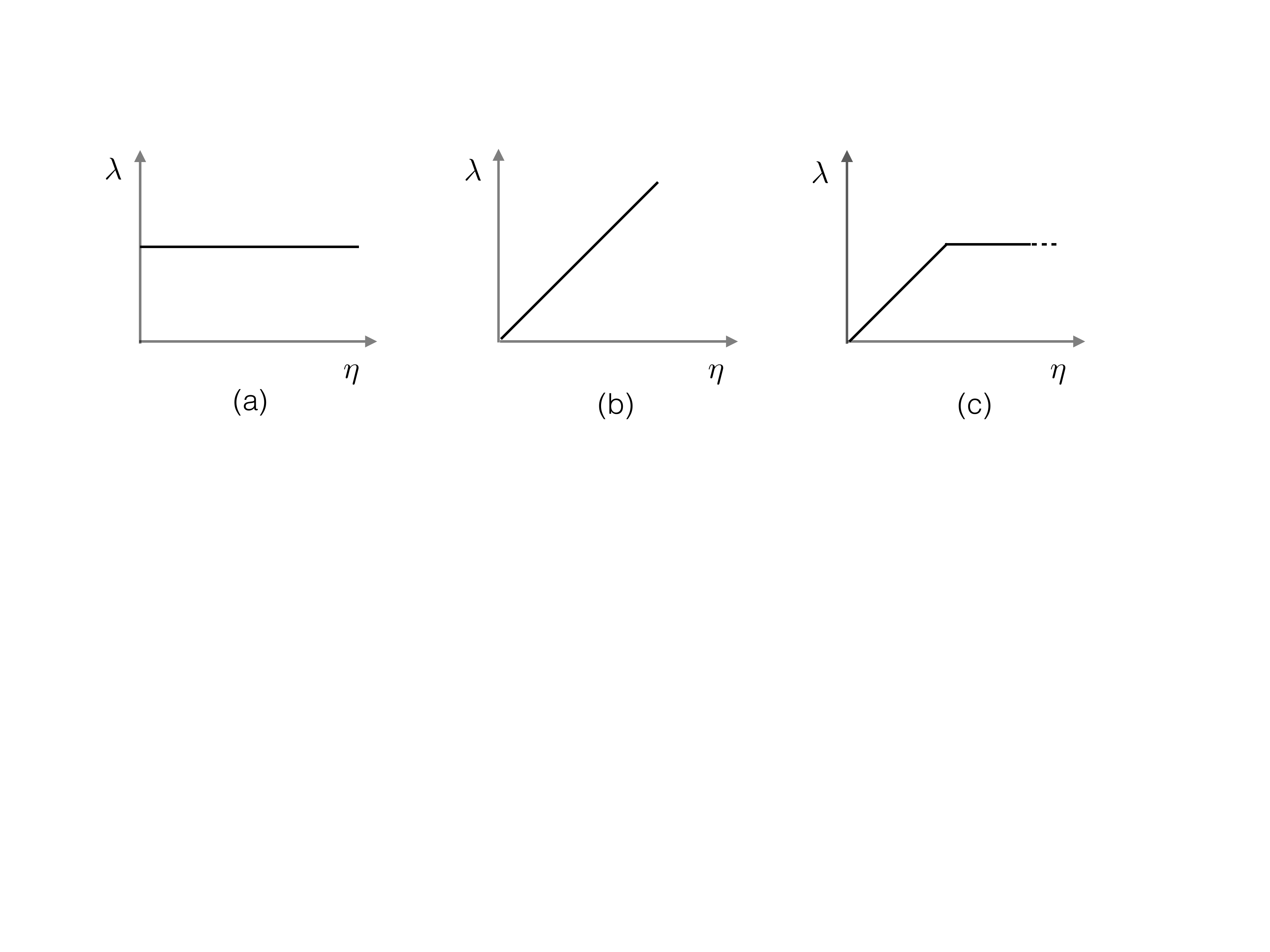}
\vspace{-6.5cm}
\caption{$\lambda(\eta)$ for the Abelian, non-Abelian and Maldacena-N\'u\~nez solutions in (a),(b),(c).}
\label{profiles}
\end{figure}

\begin{figure}[h]
\centering
\includegraphics[scale=0.3]{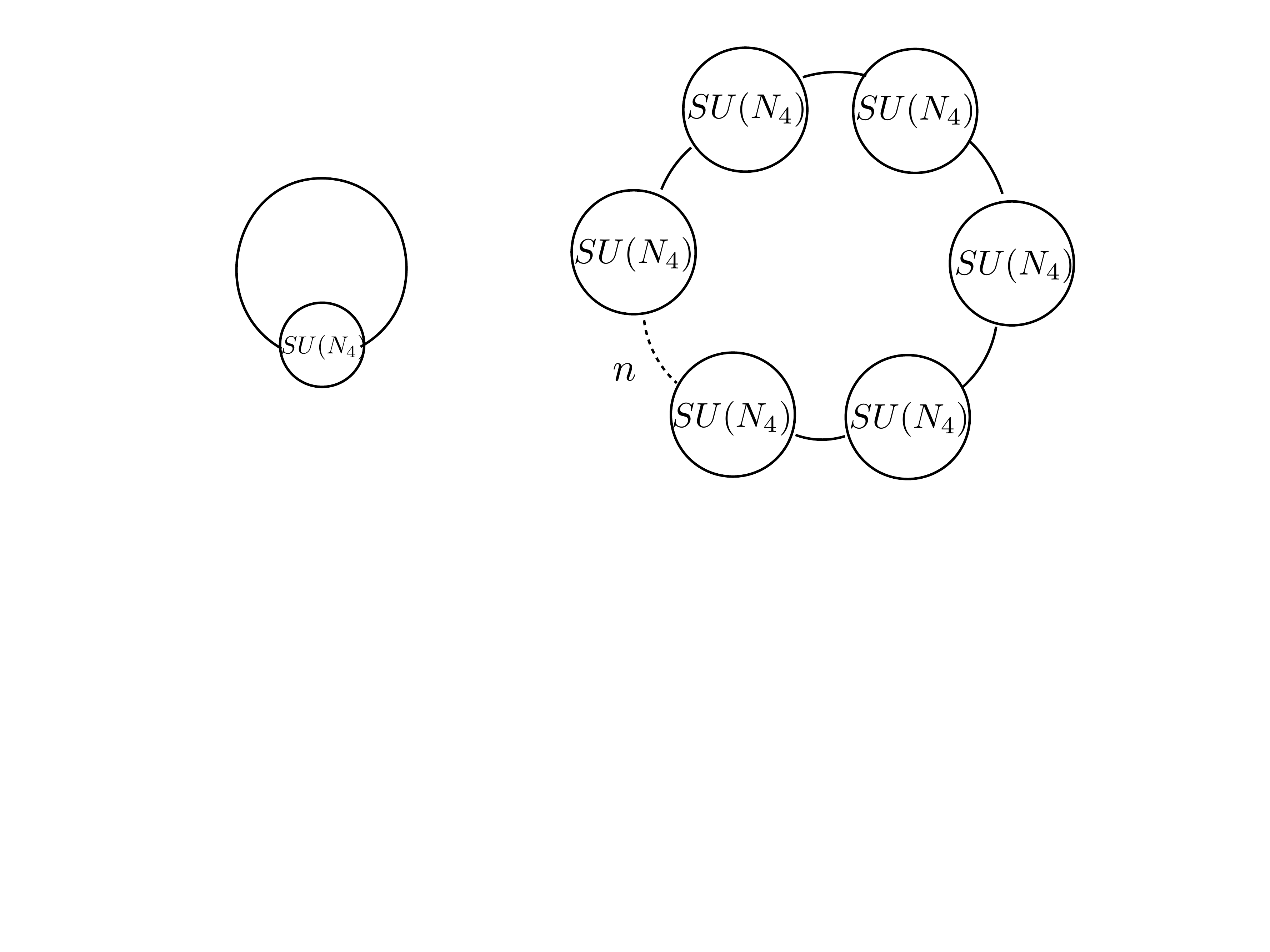}
\vspace{-4cm}
\caption{$\mathcal{N}=2$ quivers associated to $\mathcal{N}=4$ SYM and its  $\mathbb{Z}_n$ orbifold.}
\label{abelianquivers}
\end{figure}

\subsection{The case of the Non-Abelian T-dual of $AdS_5\times S^5$}

For the non-Abelian T-dual background in eq.(\ref{ads5xs5natd}), redefining variables again as in eq.(\ref{redefinition}), 
 the potential function  and charge density  are found to be 
 \cite{Sfetsos:2010uq,Macpherson:2015tka},
 \beq
V_{NATD}=\eta(\log\sigma -\frac{\sigma^2}{2})+\frac{\eta^3}{3};\;\;\;\; 
\lambda(\eta)=\eta.\label{potentialdensityNATD}
\eeq 
Notice that this background, does satisfy the condition $\lambda(0)=0$.
The charge density $\lambda(\eta)$ gives the profile (b) in Figure \ref{profiles},
which, following the rules summarised in the previous subsection, is in correspondence with a long quiver with gauge group $U(1)\times SU(2)\times SU(3)\times SU(4)\times ...$
and bifundamental hyper-multiplets connecting the nodes, as depicted in Figure \ref{linear}.
Interestingly, it describes the ``tail'' quiver  that appears in the Argyres-Seiberg dual of the $\mathcal{N}=2$ conformal quiver with $(N-1)$ $SU(N)$ gauge groups discussed
in \cite{Gaiotto:2009gz}. It is associated to a dual CFT with infinite ordinary punctures. 

This quiver is in full agreement with the brane set-up that we described in the previous section, depicted in Figure \ref{branenonab}. Indeed, rescaling 
the metric as in \cite{Gaiotto:2009gz} we find that the charge density becomes $\lambda(\eta)=\eta N_6$ and the gauge group $SU(N_{6})\times SU(2N_6)\times SU(3N_6)\times SU(4N_6)\times ...$, which is in correspondence with a configuration of $nN_6$ D4-branes stretched between NS5-branes located at $r_n=n\pi$, consistently with our analysis. As in  \cite{Gaiotto:2009gz} the elementary punctures are associated to the NS fivebranes, of which there are strictly an infinite number, for $r\in \mathbb{R}^+$. 

Next, we propose two possible ways of {\it completing} this quiver that give the right answer for the central charge (that we shall
 compute in Section \ref{QFT}). As we will
comment, completing the quiver implies an analog 'completion' in the geometry, that is, a way to effectively work with an $\eta$ or $r$-coordinate with finite range.
This is a nice example of the CFT 'informing' the dual geometry.

\begin{figure}
\centering
\includegraphics[scale=0.3]{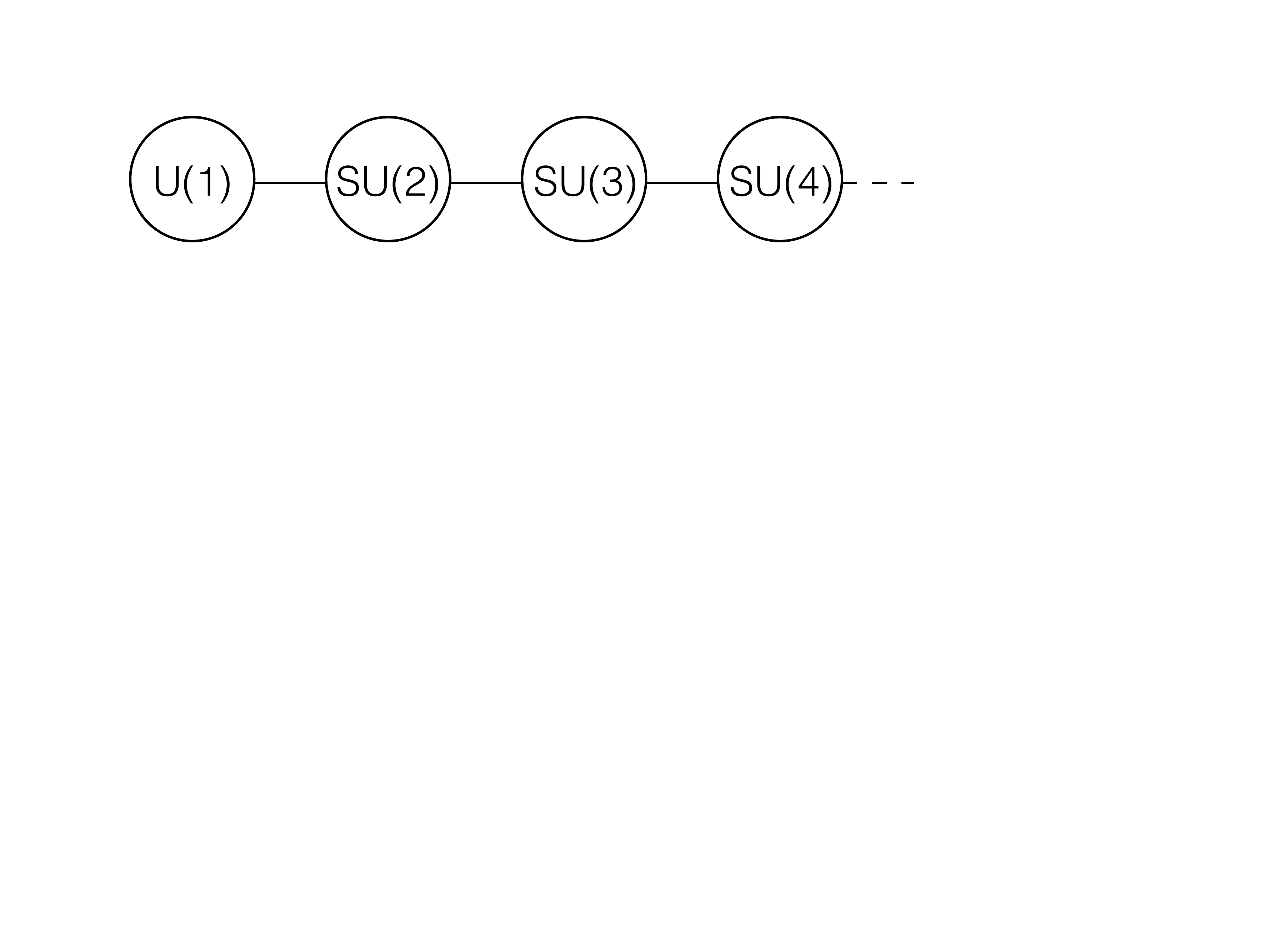}
\vspace{-6cm}
\caption{Quiver associated to the non-Abelian T-dual background.}
\label{linear}
\end{figure}

The first  {\it completion} is given by  a linear quiver consisting on  a  long chain of  
gauge groups $SU(N_{6})\times SU(2N_{6})\times SU(3N_{6})
\times....\times SU\big((p-1)N_{6}\big)$, 
finishing with a flavour group $SU(p N_{6})$ as in Figure \ref{linear2}. This corresponds to the profile (a) in Figure \ref{profiles3}. Notice that at the point $\eta=p-1$ we have a change of slope of size $pN_6$, which represents the  $SU(pN_6)$ flavour group, as summarized above.
The number of flavours and the number 
of colours are such that $N_f=2 N_{4}=2 N_c$ is satisfied for each node in the quiver, leading to 
a vanishing beta function
in correspondence with the $AdS_5$ factor in the geometry. 
This quiver can be put in correspondence with the brane set-up that we proposed in Section \ref{branerealisation}. Indeed, a long array of NS-five branes
with a given number of D4 branes, $N_4= n N_6$, in the $n$-th interval describes our conformal quiver. The flavour group that we proposed to complete the quiver can be thought of as a set of $pN_6$ semi-infinite D4-branes or D6 branes on the right of the $p^{th}$ NS5 brane. 

We stress that our addition of the flavour group should be thought of (via holography) as a way of 
ending the space or giving a finite range to the $r$-coordinate.
 The consequence of this will be reflected in the  holographic calculation of observables, that will involve integrals and 
sums in a finite range as we show in Section \ref{QFT}. 

\begin{figure}
\centering
\includegraphics[scale=0.4]{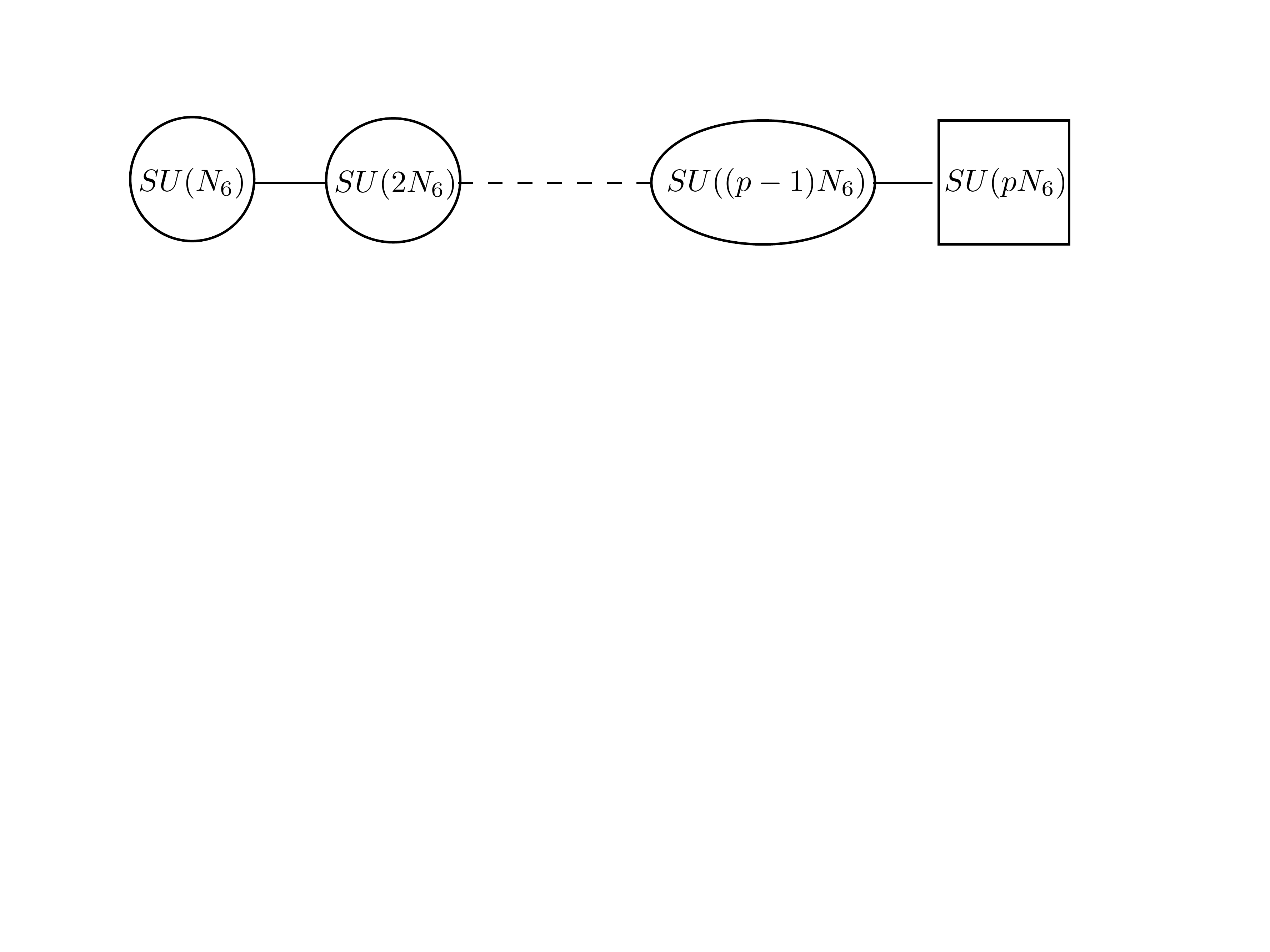}
\vspace{-8cm}
\caption{{\it Completed} non-Abelian quiver.}
\label{linear2}
\end{figure}

\begin{figure}
\centering
\includegraphics[scale=0.45]{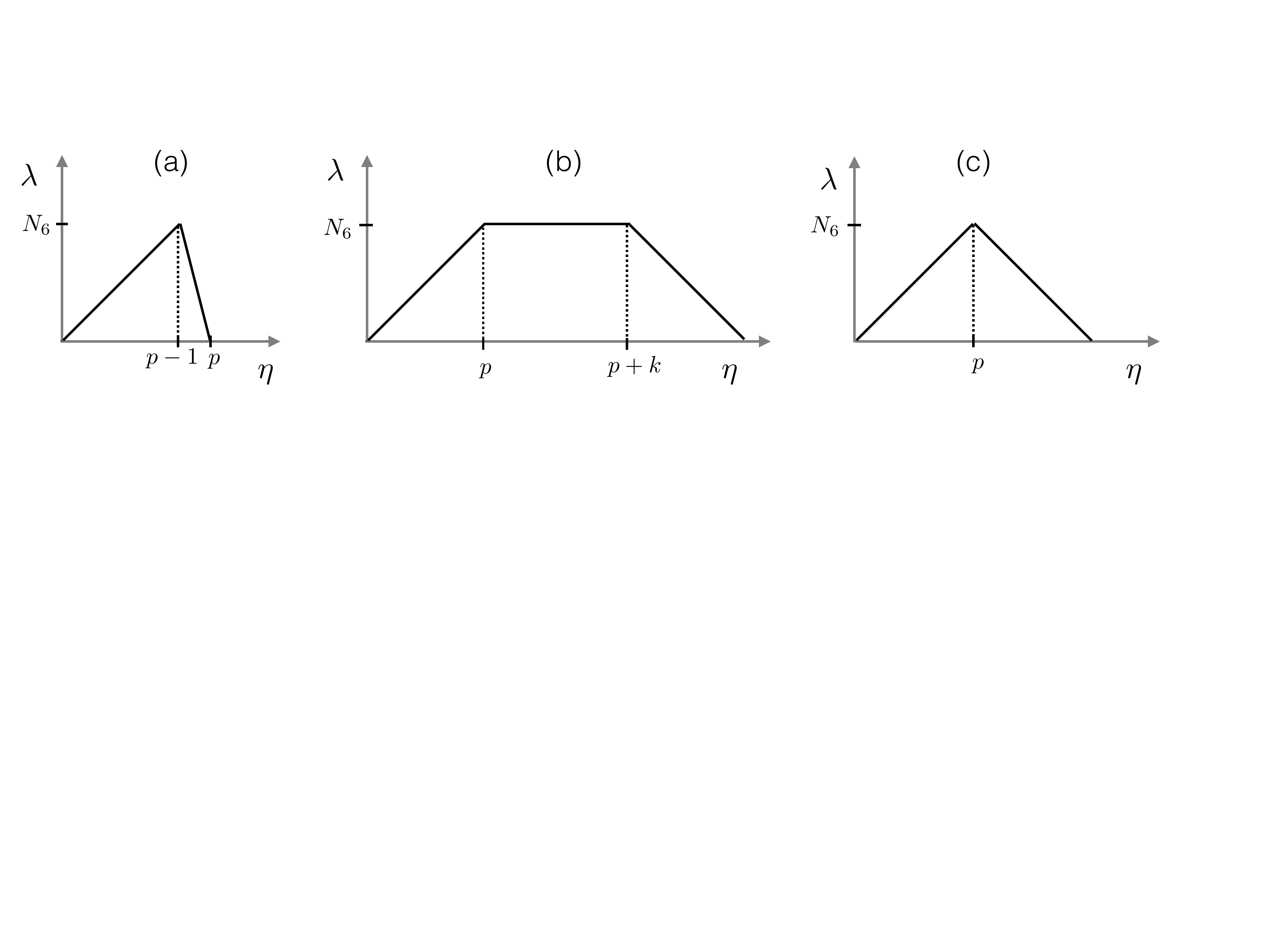}
\vspace{-7cm}
\caption{(a): $\lambda(\eta)$ for the {\it completed} quiver depicted in Fig.\ref{linear2}; (b) ((c)): $\lambda(\eta)$ for the {\it completed} quiver in Fig.\ref{completed2} for $k>1$ ($k=0$).}
\label{profiles3}
\end{figure}

We can be more precise regarding the geometry associated with this completed quiver.  Following the Gaiotto-Maldacena rules
\cite{Gaiotto:2009gz} and the content of the papers \cite{ReidEdwards:2010qs}, \cite{Aharony:2012tz}
we can write the charge density $\lambda(\eta)$
associated with it as
\[  \frac{\lambda(\eta)}{N_{6}}= \left\{
\begin{array}{ll}
      \eta & 0\leq \eta\leq p-1 \\
      (1-p)\eta+(p^2-p) & (p-1)\leq \eta\leq p \\
\end{array} 
\right. \]
As in \cite{ReidEdwards:2010qs}, the idea is to write the $\dot{V}$, and the respective charge density $\lambda(\eta)$ associated with our quiver in Figure \ref{linear2}, 
as a superposition of $\dot{V}_{MN}$ (and $\lambda_{MN}$) in eq.(\ref{vdotmn}). 
Following the treatment in
\cite{ReidEdwards:2010qs},\cite{Aharony:2012tz}, we find the precise $\dot{V}$, in terms of which (and its derivatives) the background is written,
\bea
& & \dot{V}(\sigma,\eta)=\frac{1}{2}\sum_{m=-\infty}^{\infty} 
(p-1)N_{6} \Big[\sqrt{\sigma^2 +(\eta-2m p-p)^2}-
\sqrt{\sigma^2 +(\eta-2m p+p)^2}     \Big]\nonumber\\
& & 
-p N_{6}\Big[\sqrt{\sigma^2 +(\eta-2m p-N_{6})^2}-
\sqrt{\sigma^2 +(\eta-2m p+ N_{6})^2}  \Big].
\label{potencialcompleto}
\eea
\begin{figure}
\centering
\includegraphics[scale=0.6]{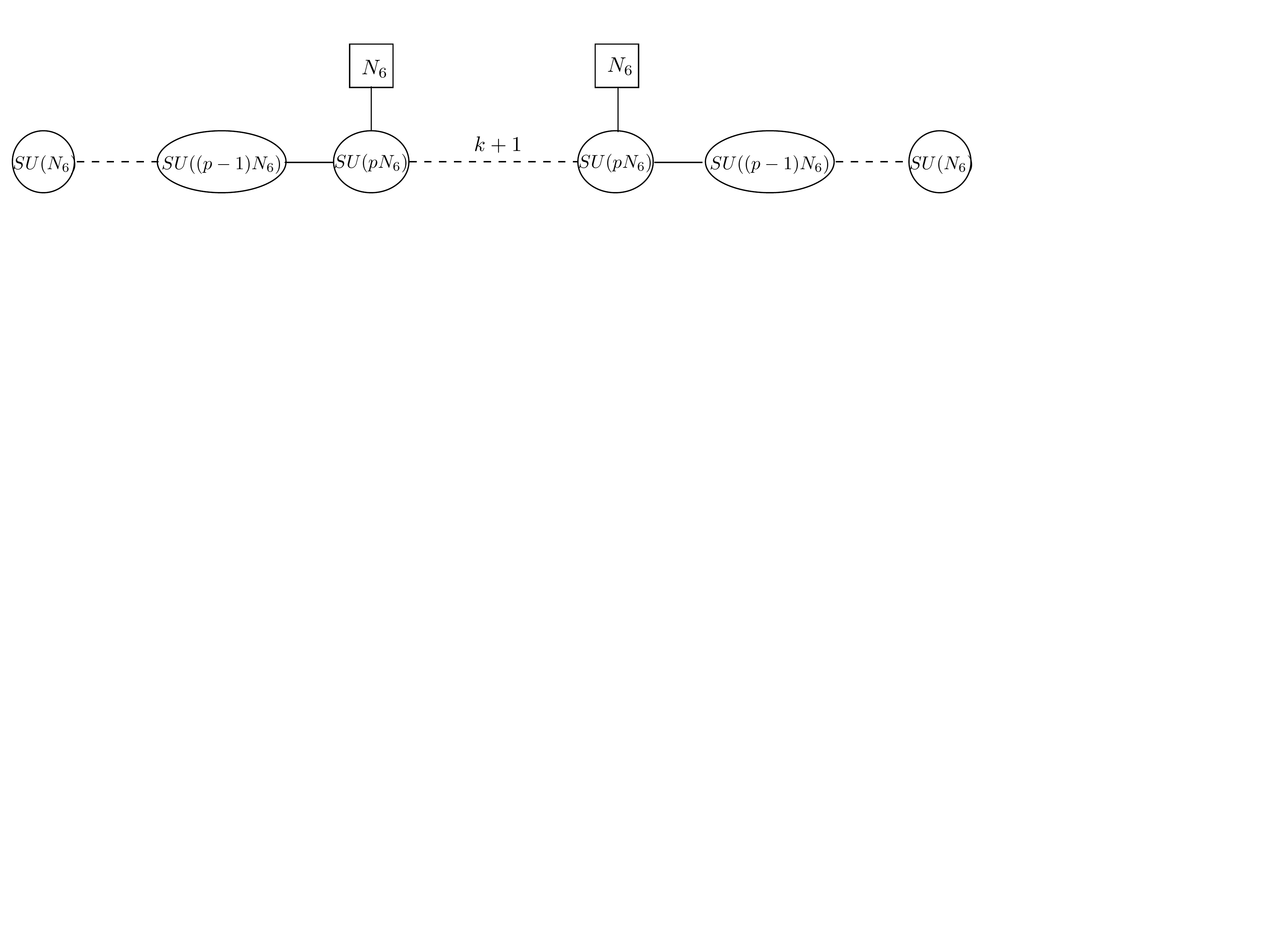}
\vspace{-13cm}
\caption{{\it Completed} non-Abelian quiver.}
\label{completed2}
\end{figure}
Again, note that at the point $\eta=p$, we have a change in the slope of $\lambda(\eta)$ of size $pN_6$. This is in correspondence
with the $SU(pN_6)$ global symmetry realised on the  D6-flavour branes, which provide the boundary condition to end the space at $\eta=p$.

A second possibility to {\it complete} the non-Abelian quiver in Figure \ref{linear} is to consider its $\mathbb{Z}_2$ orbifold  in Figure \ref{completed2}. This completed orbifold makes full use of the idea, discussed in Section \ref{seccionrelacion} and further stressed at the end of this section, that the Abelian theory functions as a sort of completion for $r\rightarrow\infty$ of the non-Abelian one.
The orbifold in Figure \ref{completed2} has associated a charge density
\[  \frac{\lambda(\eta)}{ N_{6}}= \left\{
\begin{array}{ll}
      \eta & 0\leq \eta\leq p \\
      p & p \leq \eta\leq p+k \\
      2p+k-\eta & p+k\leq \eta\leq 2p+k,
\end{array} 
\right. \]
which corresponds to the profile (b)  (or (c) for $k=0$) in Figure \ref{profiles3}. The associated $\dot{V}(\sigma,\eta)$ potential is given by
\bea
& & \dot{V}(\sigma,\eta)=\frac{N_6}{2}\sum_{m=-\infty}^{\infty}\sum_{l=1}^3 \sqrt{\sigma^2+ \big(\nu_l + 2m(2P+k) -\eta \big)^2}- \sqrt{\sigma^2 +\big(\nu_l -2m(2p+k) +\eta\big)^2},\nonumber\\
& &\nu_1=p,\;\;\; \nu_2=p+k,\;\;\; \nu_3=-(2p+k).\label{completequiver2a}
\eea

We wish to point out an interesting feature, relating the solution characterised by $\dot{V}_{MN}$ in eq.(\ref{vdotmn}), 
with the analog  
derivatives  $ \dot{V}_{ATD}$ and $\dot{V}_{NATD}$ (this relation is cleaner for  the derivatives of the potentials $\dot{V}$, but holds also without performing the $\sigma$-derivative). As it can be inferred from the charge densities displayed in Figure \ref{profiles},  $V_{MN}$
should interpolate between the Abelian (for large $\eta$) and the non-Abelian (for small $\eta$) backgrounds. This is indeed the case. Expanding $\dot{V}_{MN}$ close to
$(\sigma,\eta)\sim(0,0)$, one finds,  up to order $O(\eta^a\sigma^b)$, with $a+b<4$, that $\dot{V}_{MN}\approx \dot{V}_{NATD}$. This was already observed in 
\cite{Macpherson:2015tka}. More interestingly, one can perform an expansion for $\sigma\sim 0$ and large values of $\eta$, whose result is a $\dot{V}_{MN}\approx \dot{V}_{ATD}$--
up to the same order in the expansion as above. 

Further to this, the solution characterised by $V_{MN}$ is smooth, while both backgrounds obtained by T-duality are singular at $\sigma=1$.
A 3-d plot (in Figures \ref{figuraa},\ref{figurab}) of the three potentials shows a very good matching between $\dot{V}_{MN}$ with $\dot{V}_{NATD}$ (for small $\eta$) and between $\dot{V}_{MN}$ 
and  $\dot{V}_{ATD}$ (for large $\eta$), both for $\sigma\sim 0$. For values of $\sigma\sim 1$, $\dot{V}_{MN}$ differs from both $\dot{V}_{ATD}$ and $\dot{V}_{NATD}$, that lead to a singularity at $\sigma=1$. Hence, using the superpositions in eqs.(\ref{potencialcompleto}),  (\ref{completequiver2a}) smoothes-out the singular spaces obtained through Abelian and non-Abelian T-duality.  Indeed, notice that the summation of $\dot{V}_{MN}$ functions has two effects: on
one side, it bounds the range of the radial coordinate $\eta$ or $r$. On the
other hand, as we explained, it smoothes out the T-dual geometries at $\sigma=1$, where
the backgrounds were singular.

 \begin{figure}
\centering
\includegraphics[scale=0.6]{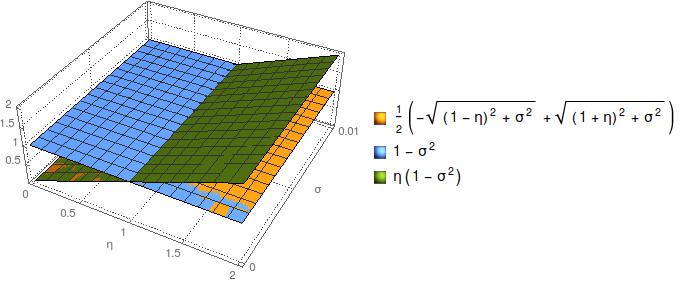}
\caption{In this Figure we see the three functions $\dot{V}_{MN}$ (in orange), $\dot{V}_{ATD}$ (in blue)
and $\dot{V}_{NATD}$ (in green) superposed. For small values of the coordinate $\sigma\sim 0$, $\dot{V}_{MN}$ approximates $\dot{V}_{NATD}$ in the interval $0\leq\eta\leq 1$, while it fits $\dot{V}_{ATD}$ for $\eta>1$.}
\label{figuraa}
\end{figure}

  \begin{figure}
\centering
\includegraphics[scale=0.6]{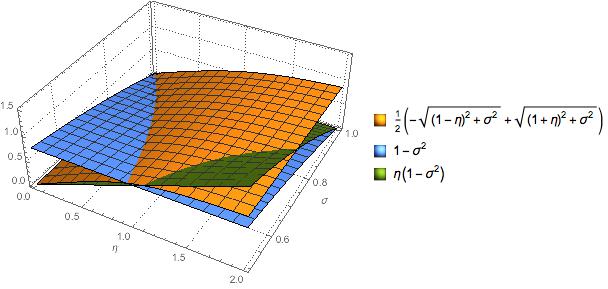}
\caption{The plot of $\dot{V}_{MN}$ (in orange), $\dot{V}_{ATD}$ (in blue)
and $\dot{V}_{NATD}$ (in green) superposed, for values of $\sigma\sim 1$.
While both $\dot{V}_{ATD},\dot{V}_{NATD}$ vanish, $\dot{V}_{MN}$
does not. This is what resolves the singularity in the backgrounds obtained by both T-dualities.}
\label{figurab}
\end{figure}

Let us discuss another interesting feature of the backgrounds we have presented. Once completed as indicated by the
associated quivers, and encoded by the functions
$\dot{V}(\sigma, \eta)$ given in eqs. (\ref{potencialcompleto})-(\ref{completequiver2a}),
these solutions correspond to the ``one NS5-stack'' class of backgrounds, according to
the classification of \cite{Aharony:2012tz}. This implies that the couplings
of each of the gauge groups are arbitrarily large. The parameters associated
with the marginal coupling of each gauge group are only visible in the ``many NS5
stacks'' backgrounds of \cite{Aharony:2012tz}.
Our solutions, obtained via a generating technique from $AdS_5\times S^5$
do not contain those parameters to begin with, so they are dual to the quivers we proposed
in the limit of very large gauge couplings (the NS5 branes have sub-stringy
separations when the $\sigma$-coordinate grows large).

To close this section we make a  couple of comments that will be useful when comparing observables of the CFT calculated with  the Abelian and non-Abelian T-dual backgrounds.
The expressions for 
the potentials and charge densities for the Abelian and non-Abelian T-duals show
that these quantities in the Abelian  
T-dual background, eq.(\ref{potentialdensityTD}), 
give place to those in the non-Abelian 
T-dual one, in eq.(\ref{potentialdensityNATD}), by superposition 
(or integration) in the $\eta$-direction. 
Intuitively, this suggests that non-Abelian T-duality can be thought 
of as a form of  'superposition' of Abelian T-dualities. 
It is also interesting to replace $n N_{6}\to N_{4}$ in each 
interval $[n\pi,(n+1)\pi]$ and see that our quiver 
dual to the non-Abelian T-dual background is identical to the quiver describing
the Abelian background with $0\leq\psi\leq n\pi$. 
The main difference is that the linear quiver dual to the background 
of eq.(\ref{ads5xs5natd}) does not close in a circle.
There is in fact more to this intuitive view, as we will stress in the next section.

In summary, inspired by the Gaiotto-Maldacena formalism and the brane realisations of 4d ${\cal N}=2$ CFTs
studied by Witten, we have proposed a particular quiver as dual to the non-Abelian $AdS_5$ solution.
We have discussed possible ways of 'completing' this quiver, read the profiles  $\lambda(\eta)$ associated to them, attained the boundary conditions and smoothed-out the backgrounds. 

We wish to check now our proposal against some field theory observables. These will be calculated in the gravitational and the CFT descriptions
of our system. To this we turn now.

\section{Observables in the Quantum Field Theory}\label{QFT}

In this section we compute important physical quantities of the 
quantum field theories associated with the quivers proposed in the previous section as duals to the Abelian and non-Abelian backgrounds. We show that they are perfectly consistent with the holographic results. 
This provides crucial complementary information
to understand the field theoretical operation associated with 
non-Abelian T-duality. We start analysing the central charges.

\subsection{Central Charges}\label{seccioncentralcharge}

We can calculate holographically the central charges of the conformal theories 
associated to the Abelian and non-Abelian geometries using the formalism developed in
 \cite{Klebanov:2007ws,Macpherson:2014eza}. Briefly and to set notation, 
 for a string-theory dual to a $(d+1)$-dimensional QFT, 
with line element 
\begin{equation}
 ds^2=a(R,\theta^i) \Big[dx_{1,d}^2 +  b(R) dR^2\Big] + g_{ij}(R,\theta^i)d\theta^i d\theta^j,  
 \end{equation}
and dilaton $\Phi(R,\theta_i)$, two quantities $\hat{V}_{int},\hat{H}$ 
can be defined,
 \begin{equation}
 \hat{V}_{int}=\int\! d\theta^i\sqrt{\det [g_{ij}] e^{-4\Phi} a^{d}},\qquad \hat{H}=\hat{V}_{int}^2,
\end{equation}
in terms of which, the central charge for the $(d+1)$-dimensional QFT reads,
\beq
c=d^d\frac{b^{d/2 } \hat{H}^{\frac{2d+1}{2}}}{G_{N,10} (\hat{H}')^d}\, .
\label{formulacentralcharge}
\eeq
Let us first compute 
the central charge for the original 
$AdS_5\times S^5$ background in eq.(\ref{ads5xs5}). In this case, we have
\beq
a(R,\theta^i)=\frac{4R^2}{L^2}, \;\; b(R)= \frac{L^4}{R^4},\;\; d=3,\;\;\;
\hat{H}= 32^2 4^3 \pi^6 L^4 R^6={\cal N}^2_{AdS5} R^6,
\label{centralads5xs5previa}
\eeq
and 
\bea
 c=\frac{4L^8}{\pi^2\alpha'^4}= \frac{N_{3}^2}{4},
\label{centralads5xs5}
\eea
where we have used the quantisation condition in eq.(\ref{quantisationd3}), the value $G_{N,10}=8\pi^5 g_s^2\alpha'^4$, and we set $g_s=1$.

\subsubsection{The  central charge for the Abelian T-dual of $AdS_5\times S^5$}

For the Abelian T-dual background in eq.(\ref{ads5xs5td}) 
we have the same values of $a(R,\theta_i),b(R)$ and $d$ as in 
$AdS_5\times S^5$. Using the quantisation condition 
in eq.(\ref{charged4}) we find,
\bea
& & \hat{H}= 2^{12}\pi^6 L^4 R^6={\cal N}^2_{AdS5ATD} R^6\nonumber\\
& & c= \frac{L^8}{\pi^2 \alpha'^4}= \frac{N_{4}^2}{4},
\label{centralads5xs5td}
\eea
which is the expected result, as the central charge is invariant 
under T-duality. Note that even if there is a rescaling by a factor of 2 in the conserved charges associated to the original and T-dual backgrounds, the central charges are the same in terms of the respective conserved charges, showing the equivalence of the associated CFTs. We would like to stress that this background provides an example of a Gaiotto-Maldacena geometry, related to M5-branes, whose central charge does not scale with $N^3$. We will elaborate more on this in Section \ref{concl}. 

For field theories 
with ${\cal N}=2$ SUSY, the central charge is written in terms 
of the number of degrees of freedom 
contained in vector multiplets, $n_v$, and the analog number
for hyper-multiplets, $n_h$, as \cite{Shapere:2008zf}, 
\beq
c=\frac{1}{12}(2n_v+n_h).\label{formulita}
\eeq
Using this, we can check that the quiver 
consisting of one 
gauge group $SU(N_{4})$ with one vector multiplet
and one adjoint hypermultiplet, depicted in Figure \ref{abelianquivers},
gives the central charge 
\beq
c=\frac{2(N_{4}^2-1) + N_{4}^2}{12},\label{util}
\eeq 
that in the limit
of large number of four branes coincides with the
holographic result in eq.(\ref{centralads5xs5td}).

In turn, if we let $\psi$ cover the $[0,\pi]$ 
interval $n$ times, that is $\psi \in [0,n\pi]$, the central charge reads  
\bea
c=n \frac{L^8}{\pi^2 \alpha'^4}=n \frac{N_{4}^2}{4},
\label{centralads5xs5tdn}
\eea
which is precisely that of the $\mathbb{Z}_n$ orbifold of $\mathcal{N}=4$ SYM with gauge group $SU(N_4)$, $c=\frac{(nN_4)^2}{4n}$, see 
\cite{Alishahiha:1999ds}.



We now check that the circular quiver in Figure \ref{abelianquivers} matches the holographic result.
We can count the number of degrees of freedom present 
in vector and hyper-multiplets,
\begin{equation}
n_v=n(N_{4}^2-1),\;\;\; n_h=n N_{4}^2,
\end{equation}
to finally obtain
\begin{equation}
c=\frac{3 n N_{4}^2-2n}{12}=n \frac{N_{4}^2}{4}(1-\frac{2}{3 N_{4}^2})\approx n \frac{N_{4}^2}{4}.\label{xxy}
\end{equation}
In agreement with eq.(\ref{centralads5xs5tdn}) and reference \cite{Alishahiha:1999ds}. Let us now study our non-Abelian T-dual system.




\subsubsection{The  central charge for the non-Abelian T-dual of $AdS_5\times S^5$}

Here, we analyse the non-Abelian T-dual case. The central charge of the geometry in eq.(\ref{ads5xs5natd}) was 
calculated in 
\cite{Macpherson:2015tka}. We have the same values 
of  $a(R,\theta_i),b(R)$ and $d$ as in the previous backgrounds. 
Using now the quantisation condition 
in eq.(\ref{chargesnatd}) and a range for  the $r$-coordinate 
between 0 and $n\pi$ we find,
\bea
& & \hat{H}=\Big[  64 L^2 \pi^2  \int_{0}^{n\pi}\!r^2dr\Big]^2 R^6
={\cal N}^2_{AdS5NATD} R^6,\nonumber\\
& & c=\frac{N_{6}^2}{4\pi^3}\int_0^{n\pi} r^2 dr=\frac{N_{6}^2 N_{5}^3}{12}
\label{centralnatdfinal}
\eea
where $N_{5}=n$ is the number of NS5-branes in the $[0,n\pi]$ interval. Interestingly, in this calculation we see the $N_5^3$ scaling with the number of NS5-branes in Type IIA (or M5-branes in the eleven dimensional lift) appearing due to the integration range in the whole $[0,n\pi]$ interval. 
We now check that the completed quivers proposed to describe the CFT dual to this background, see Figures \ref{linear2} and {\ref{completed2}}, reproduce the result of eq.(\ref{centralnatdfinal}).

For the completed quiver consisting on a long chain of gauge groups $SU(N_6)\times SU(2N_6)\times .. \times SU((p-1)N_6)$ and finishing with a flavour group $SU(pN_6)$ in Figure \ref{linear2}, we can count the number of degrees of freedom in vector multiplets and hyper-multiplets,
\bea
& & n_v=\sum_{k=1}^{p-1} k^2 N_{6}^2 -1= N_{6}^2\Big[  \frac{p^3}{3} -\frac{ p^2}{2} +\frac{p}{6} +\frac{1-p}{N_{6}^2}  \Big] ,\nonumber\\
& & n_h=\sum_{k=1}^{p-1} k(k+1)N_{6}^2=N_{6}^2 (\frac{p^3-p}{3}),
\label{zzxx}
\eea
obtaining for the central charge,
\beq
c=\frac{N_{6}^2 p^3}{12}\Big[    1-\frac{1}{p} -\frac{2}{p^2N_{6}^2} +\frac{2}{N_{6}^2p^3}\Big]\approx \frac{N_{6}^2 p^3}{12}.\nonumber
\eeq
Hence reproducing eq.(\ref{centralnatdfinal}), 
in the limit of large $N_{6}$ and large  
number of NS-five branes $p$---a limit justified when working with long quivers,  
in the approximations imposed by supergravity. 
Notice that the finite range of the $r$-integral in eq.(\ref{centralnatdfinal}) is in correspondence with the finite sum in eq.(\ref{zzxx}). This is an effect of the completion of the quiver with the flavour group at its end.

Our quiver in Figure \ref{completed2} provides in turn a completion of the infinite quiver in Figure \ref{linear} by orbifolding it by $\mathbb{Z}_2$ after adding a finite (and thus of higher order in $\frac{1}{p}$) number of $SU(pN_6)$-nodes. Accordingly it should have associated a central charge:
\begin{equation}
\label{orbifold}
c=\frac{(2N_6)^2 p^3}{24}+O(\frac{1}{p})\, .
\end{equation}
Indeed, we find that,
\bea
n_v&=&2 \big(\sum_{j=1}^{p-1}j^2N_6^2 -1 \big) + k(p^2N_6^2 -1)= \frac{2}{3} N_6^2p^3 +N_6^2p^2 (k-1) +\frac{N_6^2 p}{3} +2-2p-k,\nonumber\\
n_h&=& 2\sum_{j=1}^{p-1} j(j+1)N_6^2  +(k-1)p^2N_6^2 + 2pN_6^2 =\frac{2}{3}N_6^2p^3 +N_6^2p^2 (k-1) -\frac{2}{3}N_6^2p +2 p N_6^2 ,\nonumber
\eea
and thus
\begin{equation}
c=\frac{2n_v+n_h}{12}\approx \frac{N_6^2p^3}{6}+O(\frac{1}{p})\, ,
\end{equation}
in agreement with (\ref{orbifold}).
 
There is a third possibility to recover the right value for the central charge in 
eq.(\ref{centralnatdfinal}), in terms of `Abelian' quivers. It corresponds to the quiver depicted in Figure \ref{completed3}.
This quiver starts with a flavour group $SU(pN_6)$, followed by a long string of $\frac{p}{3}$ $SU(pN_6)$ gauge groups, finishing with another flavour group $SU(pN_6)$.
Following  \cite{ReidEdwards:2010qs},\cite{Aharony:2012tz}, we find  that the charge density $\lambda(\eta)$ and 
$\dot{V}$ for the background dual to this CFT are,
\[  \frac{\lambda(\eta)}{p N_{6}}= \left\{
\begin{array}{ll}
      \eta & 0\leq \eta\leq 1 \\
      1 & 1 \leq \eta\leq \frac{p}{3} \\
      1+\frac{p}{3}-\eta & \frac{p}{3}\leq \eta\leq 1+\frac{p}{3}
\end{array} 
\right. \]
and 
\bea
& & \dot{V}(\sigma,\eta)=\frac{N_6}{2}\sum_{m=-\infty}^{\infty}\sum_{l=1}^3 \sqrt{\sigma^2+ \big(\nu_l + 2m(\frac{p}{3} +1) -\eta \big)^2}- \sqrt{\sigma^2 +\big(\nu_l -2m(1+\frac{p}{3}) +\eta\big)^2},\nonumber\\
& &\nu_1=1,\;\;\; \nu_2=\frac{p}{3},\;\;\; \nu_3=-1-\frac{p}{3}.\label{completequiver2}
\eea
The number of vector multiplets, hypermultiplets and central charge are
\bea
& & n_v= (p^2 N_6^2-1)\frac{p}{3}, \;\;\;\; n_h=p^2 N_6^2 (1+\frac{p}{3}).\nonumber\\
& & c=\frac{2n_v+n_h}{12}= \frac{N_{6}^2 p^3}{12}(1+\frac{1}{p}-\frac{2}{3N_6^2 p^2})\approx \frac{N_{6}^2 p^3}{12},
\eea
thus matching the result in eq.(\ref{centralnatdfinal}). 

\begin{figure}
\centering
\includegraphics[scale=0.4]{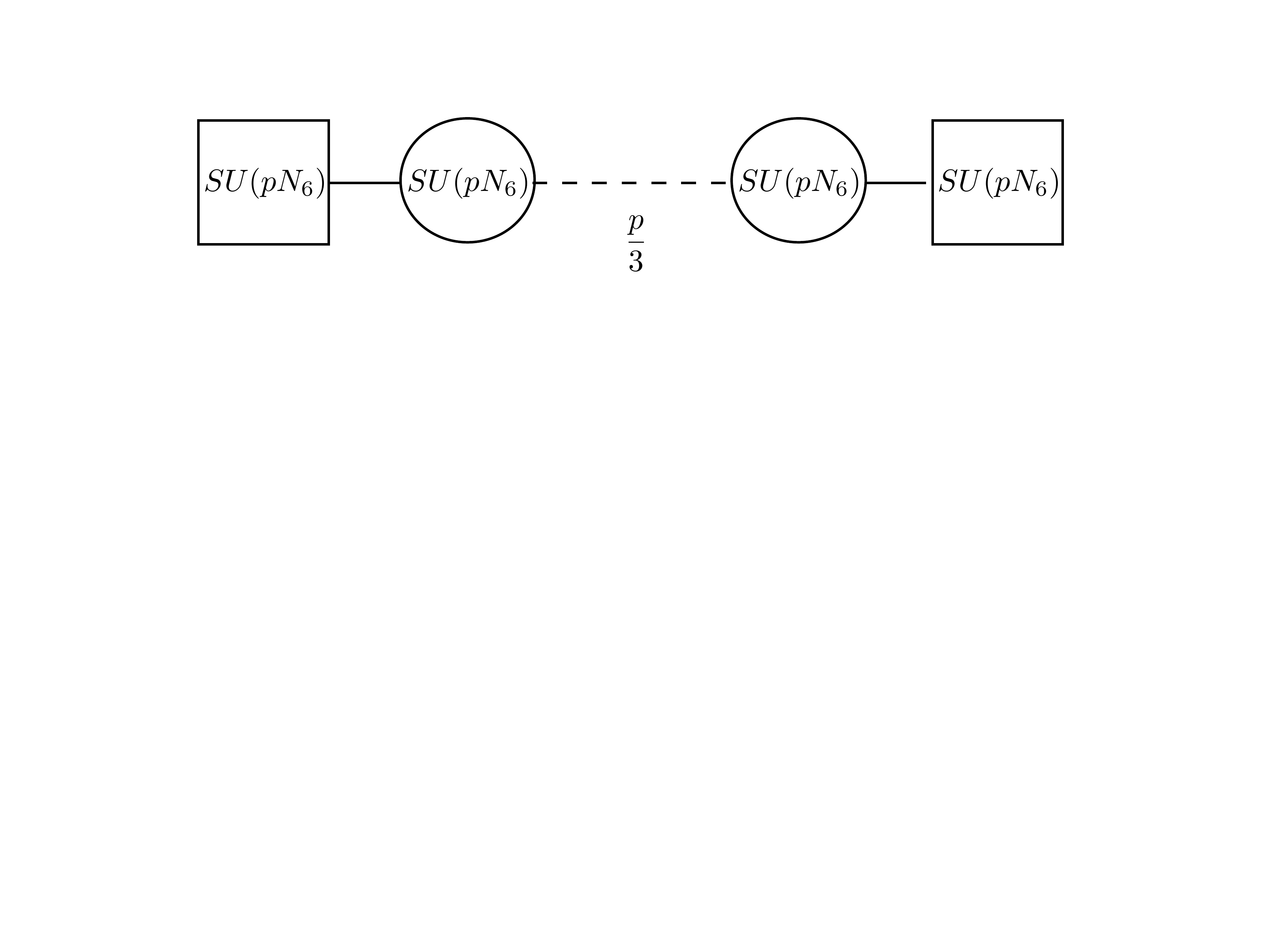}
\vspace{-8cm}
\caption{{\it Completed} non-Abelian quiver.}
\label{completed3}
\end{figure}

Finally, we wish to discuss a possible strong coupling realisation of the central charge obtained in 
eq.(\ref{centralnatdfinal}), that tries to highlight the fact that it seems to be the result of an orbifold by $\mathbb{Z}_{N_6}$ of a theory of $n$ NS5-branes:
\begin{equation}
c=\frac{N_6^2 n^3}{12}=\frac{(N_6 n)^3}{12 N_6}.
\end{equation}
To be more precise, using as building block  the $T_n$ in \cite{Gaiotto:2009we,Gaiotto:2009gz}, depicted in Fig.\ref{Tn1}, which has associated a number of vector multiplets and hypermultiplets given by 
\begin{eqnarray}
&&n_{v}=2(\frac{2n^3}{3}-\frac{3n^2}{2}-\frac{n}{6}+1) +3(n^2 -1)\\
&&n_{h}=2(\frac{2n^3}{3}-\frac{2n}{3}),
\end{eqnarray}
one obtains a central charge  \cite{Gaiotto:2009gz}
 \begin{equation}
 \label{cstrong}
 c=\frac{n^3}{3}+O(\frac{1}{n}).
 \end{equation}
 Orbifolding now by $\mathbb{Z}_{N_6/2}$, as depicted in Fig.\ref{Tn2}, we recover  $N_6$ $SU(n)$ nodes and a central charge 
\begin{equation}
c=\frac{(nN_6/2)^3}{3N_6/2}=\frac{n^3 N_6^2}{12},
\end{equation}
as in eq.(\ref{centralnatdfinal}). Note that a configuration of NS5-branes stretched between D6-branes is related by a chain of T-S-T dualities to a D4/NS5 brane set-up. Our NS5-branes would lie in the $(\mathbb{R}^{1,3},\alpha,\beta)$ directions and would be stretched between D6-branes lying on $(\mathbb{R}^{1,3},r,\chi,\xi)$.
The S-duality operation involved in this relation would imply that this configuration would be strongly coupled, which could be in correspondence with this description in terms of strongly coupled $T_n$ building blocks.

\begin{figure}[!ht]
\centering
\includegraphics[scale=0.35]{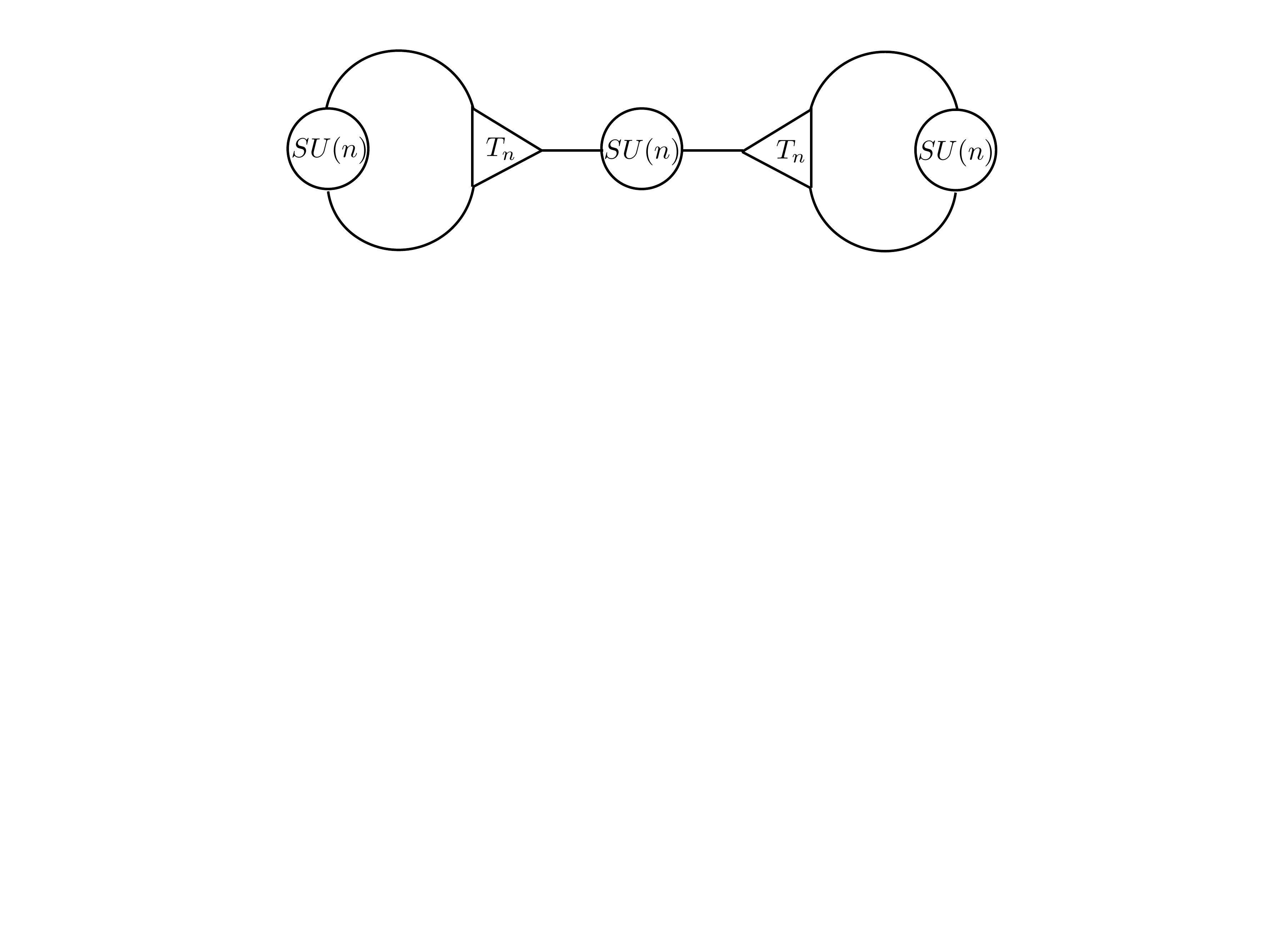}
\vspace{-7cm}
\caption{$T_n$ building block. The $SU(n)$ global symmetries are gauged.}
\label{Tn1}
\end{figure}

 \begin{figure}
\centering
\includegraphics[scale=0.4]{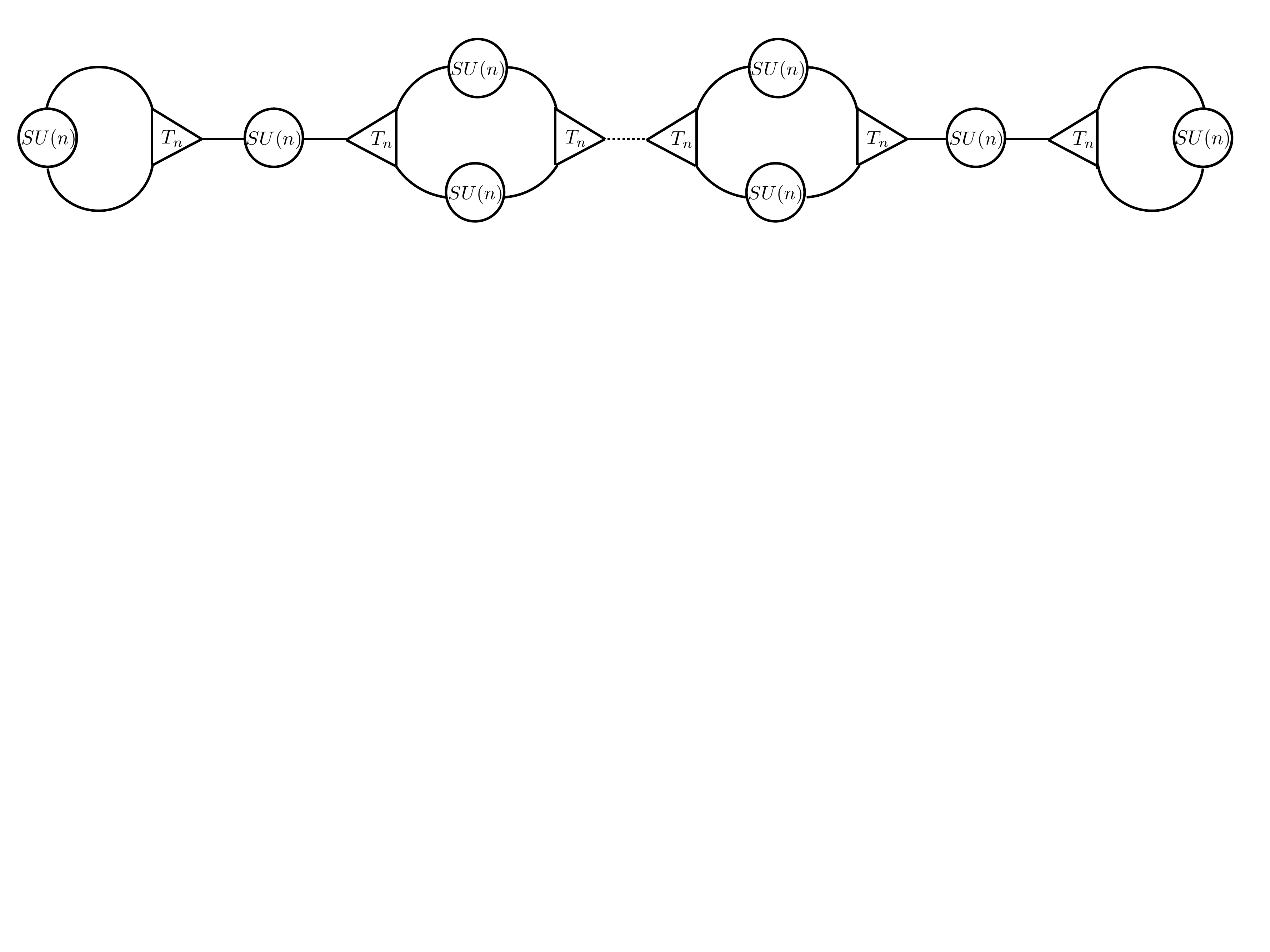}
\vspace{-8cm}
\caption{$\mathbb{Z}_{N_6/2}$ `orbifolding' of the basic $T_n$ building block.}
\label{Tn2}
\end{figure}

 These different CFTs we have proposed here can be distinguished by
either sub-leading corrections in $\frac{1}{N},\frac{1}{P}$ or by
other observables calculated in the Supergravity approximation.


\subsubsection{Relation between Abelian and non-Abelian T-duals}

Let us now check how the
relation between Abelian and non-Abelian T-dual backgrounds previously discussed, is satisfied in the field theory side.
In light of our previous discussion, we should be able to 
relate eq.(\ref{centralads5xs5td}) with eq.(\ref{centralnatdfinal}).
For this we compute the central charge as in 
eq.(\ref{centralnatdfinal}) but with $r$ varying in the 
$[n\pi,(n+1)\pi]$ interval. We obtain
\beq
c=\frac{N_{6}^2}{12}(3 n^2+3n+1)= 
\frac{N_{4}^2}{4}(1+\frac{1}{n}+\frac{1}{3n^2}),
\label{relationc}
\eeq
where we have used that in this interval $N_{4}=n N_{6}$. 
The Abelian limit of this expression amounts to 
taking $n\to\infty$ ---in correspondence with the large $r$ 
limit discussed in Section 2.3. 
In this limit eq.(\ref{relationc}) 
indeed reduces to eq.(\ref{centralads5xs5td}). 

This suggests that the CFT dual to the non-Abelian solution in the $r\in
[0,(n-1)\pi]$ interval, which, as already mentioned, contains a vector
multiplet and a bifundamental of $SU(kN_6)$ in each $[k\pi, (k+1)\pi]$ interval, for $k=0,\dots,(n-2)$, behaves effectively in the limit $n\rightarrow \infty$, as a hypermultiplet in the adjoint
of $SU(nN_6)$.

Moreover, this calculation also suggests that non-Abelian 
T-duality in an interval of length $\pi$ is a 'corrected' (that is, capturing $O(\frac{1}{n})$-finite size effects) 
version of Abelian T-duality. 

A further interesting observation is 
that the Abelian T-dual central charge $c= \frac{N_{4}^2}{4}$ 
arises as the derivative with respect to $n$ of the non-Abelian 
central charge in the $[0,n\pi]$ interval, 
given in eq. (\ref{centralnatdfinal}), recall that here $N_{5}=n$. 
This happens because we can calculate the derivative of $c$ as 
the limit when $n$ goes to infinity of the difference of its values 
in the $[0,(n+1)\pi]$ and $[0,n\pi]$ intervals, which is exactly what 
is done to obtain  eq.(\ref{relationc}). Again, 
non-Abelian T-duality seems to arise as a sort of superposition of 
Abelian T-dualities, as we already observed when we computed the 
potentials and charge densities associated to both solutions.

Another observable that behaves similarly to the central charge is the entanglement entropy, that  we study next.
\subsection{Entanglement entropy}
A very similar behaviour to that of the central charge is found for 
the entanglement entropy. Indeed, considering 
a strip geometry of size $l$ and after regularisation 
(see \cite{Kol:2014nqa} for the precise general expressions),
 we have that the entanglement entropy  and the size of the strip 
for the different four dimensional CFTs is given by,
\bea
& & \frac{2 G_{N,10}}{V_2} S_{EE}(R_*)={\cal N} L^2 \Big[ \int_{R_*}^{\infty} \frac{R^4}{\sqrt{R^6-R_*^6}} dR      -\int_{R_\Lambda}^\infty dR R  \Big]= {\cal N}L^2 R_*^2 \Big[  \int_1^\infty\frac{x^4 dx}{\sqrt{x^6-1}} -\int_0^\infty x dx \Big].\nonumber\\
& &  l(R_*)=2R_*^3 L^2\int_{R_*}^\infty \frac{dR}{\sqrt{R^4(R^6-R_*^6)}}= \frac{2L^2\sqrt{\pi} \Gamma(\frac{2}{3})}{\Gamma(\frac{1}{6})} \frac{1}{R_*}.\nonumber
\eea
The information about the CFT in question is encoded  
in the different values of ${\cal N}$ calculated for  each of the backgrounds. 
These quantities can be read from eqs.(\ref{centralads5xs5previa}),
(\ref{centralads5xs5td}), (\ref{centralnatdfinal}).
Inverting the relation $l(R_*)$ we can write, 
\bea
 & & \frac{S_{EE}(l)}{V_2}=\frac{2 \pi  \mu (\Gamma(\frac{2}{3}))^2}{G_{N,10}(\Gamma(\frac{1}{6}))^2 l^2} {\cal N} L^6, \;\;\; \mu=\int_1^\infty\frac{x^4 dx}{\sqrt{x^6-1}} -\int_0^\infty x dx .\nonumber
 \eea
It is useful to calculate the quotients
\bea
 \frac{L^6 {\cal N}_{AdS5}}{\alpha'^4}= 16 \pi^5 N_{3}^2,\;\;\frac{L^6 {\cal N}_{AdS5ATD}}{\alpha'^4}=  16  \pi^5 { N_{4}^2} ,\;\;
 \frac{L^6 {\cal N}_{AdS5NATD}}{\alpha'^4}= 16 \pi^2 {N_{6}^2} \int_0^{n\pi} r^2 dr.\nonumber
\eea
A very similar argument to that explained with 
the central charge around eq.(\ref{relationc}) 
can be made for the entanglement entropy using the results above. 
Indeed, using that $G_{N,10}= 8\pi^5 g_s^2\alpha'^4$ and setting $g_s=1$ as above, we have that the entanglement entropy per unit volume follows an area-law
in each of the CFTs, with theory-dependent coefficients,

\[  \frac{S_{EE}(l)}{V_2}= \frac{4\pi\mu}{l^2}\Big(\frac{\Gamma(\frac{2}{3})}{\Gamma(\frac{1}{6})}\Big)^2 \left\{
\begin{array}{ll}
      N_{3}^2 & AdS_5\times S^5 \\
      N_{4}^2 & {\rm ATD} \\
      \frac{N_{6}^2 N_{5}^3}{3}& {\rm NATD}.
\end{array} 
\right. \]
Once again,  we can make a correspondence between the entanglement entropy of the  
Abelian T-dual and that of the non-Abelian one, by considering the very last calculation in the interval $[n\pi,(n+1)\pi]$ followed by the large $n$ limit.
 
\subsection{Couplings}

In this section we calculate the couplings associated to the 4d CFTs dual to both the Abelian and non-Abelian T-dual backgrounds. To do so 
we switch on an electromagnetic field on a colour brane probe in these theories and compute the value of the coupling of the $F_{\mu\nu}^2$-operator. For the Abelian T-dual
background, we consider a BPS probe  D4-brane wrapped on $\psi$. For the non-Abelian T-dual we take a probe D4-brane wrapped on $r$. In this case, we find that  the D4-probes are BPS only when located at the singularity $2\alpha=\pi$. This is consistent with the fact that D4-branes have associated a non-vanishing charge $N_{4}=nN_{6}$ only in the presence of large gauge transformations, and these are defined in a 2-cycle that sits at the singularity. 


First, to set notation, we compute the gauge coupling of the original $AdS_5\times S^5$ background. In this case we consider a probe D3 brane, whose world-volume is on $\mathbb{R}^{1,3}$--the Minkowski directions-- and stands at some fixed value of the  $R$-coordinate.
The RR potential $C_4$ that follows from eq.(\ref{ads5xs5}) is $C_4=\frac{16 R^4}{L^4}dt\wedge dx_1\wedge dx_3\wedge dx_3$. Switching on an electromagnetic field on the brane (that for simplicity we take to be $F_2=F_{tx} dt\wedge dx_1$) we obtain the BIWZ-action,
\beq
-S_{BIWZ}=T_{D3} \int\! d^4 x\!\Big[ e^{-\Phi}\sqrt{\det[g+2\pi\alpha'F]} -C_4\Big]= 
T_{D3}\int\! d^4 x \,(\frac{4R^2}{L^2})^2\!\Big[  \sqrt{1-4\pi^2\alpha'^2 (\frac{L^2}{4R^2})^2 F_{tx}^2}   -1\Big].\label{biwz}
\eeq
When the combination $4\pi^2\alpha'^2 F_{tx}^2$ is small, we can Taylor expand and find that the effective Maxwell coupling is,
\bea
 S_{BIWZ}\approx -T_{D3}2\pi^2\alpha'^2 \int d^4x F_{tx}F^{tx}= -\frac{1}{4g_{D3}^2}\int d^4x F_{tx}F^{tx}\to \frac{1}{g_{D3}^2}=\frac{1}{\pi}.\label{couplingd3}
\eea
We have raised indexes with the Minkowski metric and  used that for a $Dp$-brane $(2\pi)^p g_s (\alpha')^{\frac{p+1}{2}} T_{Dp}=1$. We also set $g_s=1$ as  above. 

Let us now move to the T-dual examples. In the Abelian background of eq.(\ref{ads5xs5td}), we consider the motion of a probe D4-brane that extends in the Minkowski  and 
$\psi$-directions, for a fixed value of the $R$-coordinate. We also switch on  an electromagnetic field $F_{tx}$. We find that the relevant RR potential is in this case, $
C_5=16\sqrt{\alpha' }\frac{R^4}{L^4}dt\wedge dx_1\wedge dx_2\wedge dx_3\wedge d\psi$.
The BIWZ action takes a simple expression, after a cancellation between the dilaton and the $g_{\psi\psi}$ component of the metric,
\beq
-S_{BIWZ}=T_{D4}(\frac{4R^2}{L^2})^2\sqrt{\alpha'}\int d^5x \Big[  \sqrt{1-4\pi^2\alpha'^2 (\frac{L^2}{4R^2})^2 F_{tx}^2}   -1\Big].\label{BIWZd4}
\eeq
Expanding  for small values of $\alpha'F_{tx}$, and using a range for the $\psi$-coordinate in $[0,n\pi]$, we find
\bea
 \frac{2\pi^2}{g_{D4,A}^2}=\int_0^{n\pi} d\psi \to\frac{1}{g_{D4,A}^2}=\frac{n}{2\pi}.\label{couplingd4}
\eea
Hence, the larger the number of nodes $n$ in the quiver, the weaker this coupling becomes. For the Abelian T-dual of $AdS_5\times S^5$  ($n=1$)  this gives
\begin{equation}
\label{abeliancoupling}
\frac{1}{g_{D4,A}^2}=\frac{1}{2\pi}=\frac{N_4 \alpha'^2}{4L^4},
\end{equation}
in terms of the quantised charge $N_4$ in (\ref{charged4}). The usefulness of writing this coupling in terms of $N_4$ will be justified below.


We now move to the more interesting calculations in  the non-Abelian T-dual background of eq.(\ref{ads5xs5natd}). 
The RR potential that couples to the D4 brane in this background is,
\beq
C_5=16 \sqrt{\alpha'} (\frac{R}{L})^4 r dt \wedge dx_1\wedge dx_2\wedge dx_3\wedge dr.
\label{ppqq}
\eeq
Let us consider the case of D4 branes that extend in $\mathbb{R}^{1,3}\times r$, as in the Abelian case.
The result for the BIWZ action is
\beq
-S_{BIWZ}=T_{D4}\int d^5 x \frac{16R^4\sqrt{\alpha'}}{L^4} r\Big[ (1+\frac{L^4}{\alpha'^2r^2}\cos^4\alpha)^{1/2} \sqrt{1-4\pi^2\alpha'^2 (\frac{L^2}{4R^2})^2 F_{tx}^2}
- 1\Big].\label{biwznatd4}
\eeq
The presence of the factor $\big(1+\frac{L^4}{\alpha'^2r^2}\cos^4\alpha\big)$---coming 
from the dilaton--indicates that this brane does not preserve SUSY unless $\cos\alpha=0$. This is consistent with the fact that these branes have only associated a quantised charge in the presence of large gauge transformations, defined in non-trivial 2-cycles that must sit at the singularity $2\alpha=\pi$. Locating the brane at this point and 
expanding for small  values 
of the electromagnetic field  $\alpha' F_{tx}$, we find
\bea
\frac{2\pi^2}{g_{D4,NA}^2}=\int_{0} ^{n\pi} r dr= \frac{n^2\pi^2}{2}.
\label{couplingnatd4}
\eea
To make the correspondence between the couplings in the Abelian and the non-Abelian backgrounds, we  calculate the coupling $g_{D4,NA}$ in the interval $[n\pi,(n+1)\pi]$.
The result is,
\begin{equation}
\label{nonabcou}
\frac{1}{g_{D4,NA}^2}=\frac14 (2n+1)=\frac{n}{2} +O(1/n)=\frac{N_4 \alpha'^2}{4L^4} + O(1/n),
\end{equation}
with $N_4$ the non-Abelian D4 brane charge, given by $N_4=n N_6$, with $N_6$ as in 
(\ref{chargesnatd}). 
This gives us the gauge coupling for each of the $SU(nN_6)$ gauge groups in the non-Abelian quiver. 
It clearly shows that the coupling decreases as we approach the `Abelian' region, $n\rightarrow\infty$, leaving a strongly coupled theory in the $[0,\pi]$ interval.
Note that eq.(\ref{nonabcou}) fully agrees with the Abelian result, given in eq. (\ref{abeliancoupling}), in the large $n$ limit.  This calculation suggests, once more, that non-Abelian T-duality in a length $\pi$ interval captures $O(1/n)$-finite size effects of its Abelian counterpart.
Note as well that the observables associated with these gauge couplings, namely, the 't Hooft couplings, satisfy
\bea
\lambda_{A}=g_{D4,A}^2 N_{4,A}= 2\pi N_{4,A}.\;\;\;\;\; \lambda_{NA} =g_{D4,NA}^2 N_{4,NA}= 2\pi N_{4,A},
\eea
(here we have made explicit that in each case we have to multiply by the respective quantised charge) and so both 't Hooft coupling are exactly the same in both theories.

Finally, in order to complete this picture, we show that it is possible to construct BPS D6-branes in the non-Abelian T-dual background that can be used as colour branes in the $r\in [0,\pi]$ interval, where there are no large gauge transformations and thus no D4-branes present. As we have shown in the previous analysis of the couplings, this region corresponds to a strongly coupled region in which our description in terms of D4-branes stretched between NS5-branes may not apply.
It may be possible that a suitable description of the theory in this region is in terms of the $T_n$ building blocks in Figure \ref{Tn2}
that we have discussed.

\noindent The RR potentials that couple to the D6 branes in this background are
\begin{equation}
\label{C7}
C_7-C_5\wedge B_2=\frac{16R^4}{L^4}\alpha'^{3/2} r^2 \sin\chi dx_{1,3}\wedge dr\wedge d\chi\wedge d\xi.
\end{equation}
For the BI part of the Action, we find (setting $F_{tx}=0$ in this calculation),
\begin{equation}
\label{bid6}
S_{BI,NATD}=T_{D6}\alpha'^{3/2} \int d\xi d\chi \sin\chi \int dr r^2  \int d^4  x(\frac{4R^2}{L^2})^2 
\end{equation}
which clearly cancels against the integrated WZ contribution in eq.(\ref{C7}). Note that these branes need not be placed at the $2\alpha=\pi$ singularity in order to be BPS. This  is consistent with the fact that they exist even in the case of vanishing large gauge transformations.

\subsection{Relation with Deconstruction}\label{secciondeconstruction}

The expressions for the couplings of the Abelian and non-Abelian T-dual theories are very suggestive of deconstructed extra dimensions.  

In the spirit of \cite{ArkaniHamed:2001ca,ArkaniHamed:2001ie} a 4d $\mathcal{N}=2$ CFT can be seen as the UV completion, at distances shorter than the lattice spacing, of a latticised 5d $\mathcal{N}=2$ CFT compactified on a circle. The connection between these 4d and 5d theories is made concrete through the study of a 4d $\mathcal{N}=2$ CFT described by a circular quiver, depicted in Figure \ref{circular}, with $k$ $SU(N)$ gauge groups, all with gauge coupling $G$,  connected by bi-fundamentals. At low energies compared to the energy scale set by the expectation values of the bifundamental fields, $\Phi I$, the action describes a 5d ${\mathcal N}=2$ field theory with a latticised extra dimension with lattice spacing $a=1/(G\Phi)$, and 5d gauge coupling $g_5^2=G/\Phi$ 
 \cite{ArkaniHamed:2001ie}.  The radius of the fifth dimension is given by $2\pi R_5=k a$. For $E<<1/R_5$ the theory reduces instead to a 4d ${\mathcal N}=4$ CFT, with gauge group $SU(N)$ and gauge coupling 
\begin{equation}
\label{gSYM}
g_4^2=\frac{G^2}{k}=\frac{g_5^2}{2\pi R_5}.
\end{equation}

\begin{figure}
\centering
\includegraphics[scale=0.3]{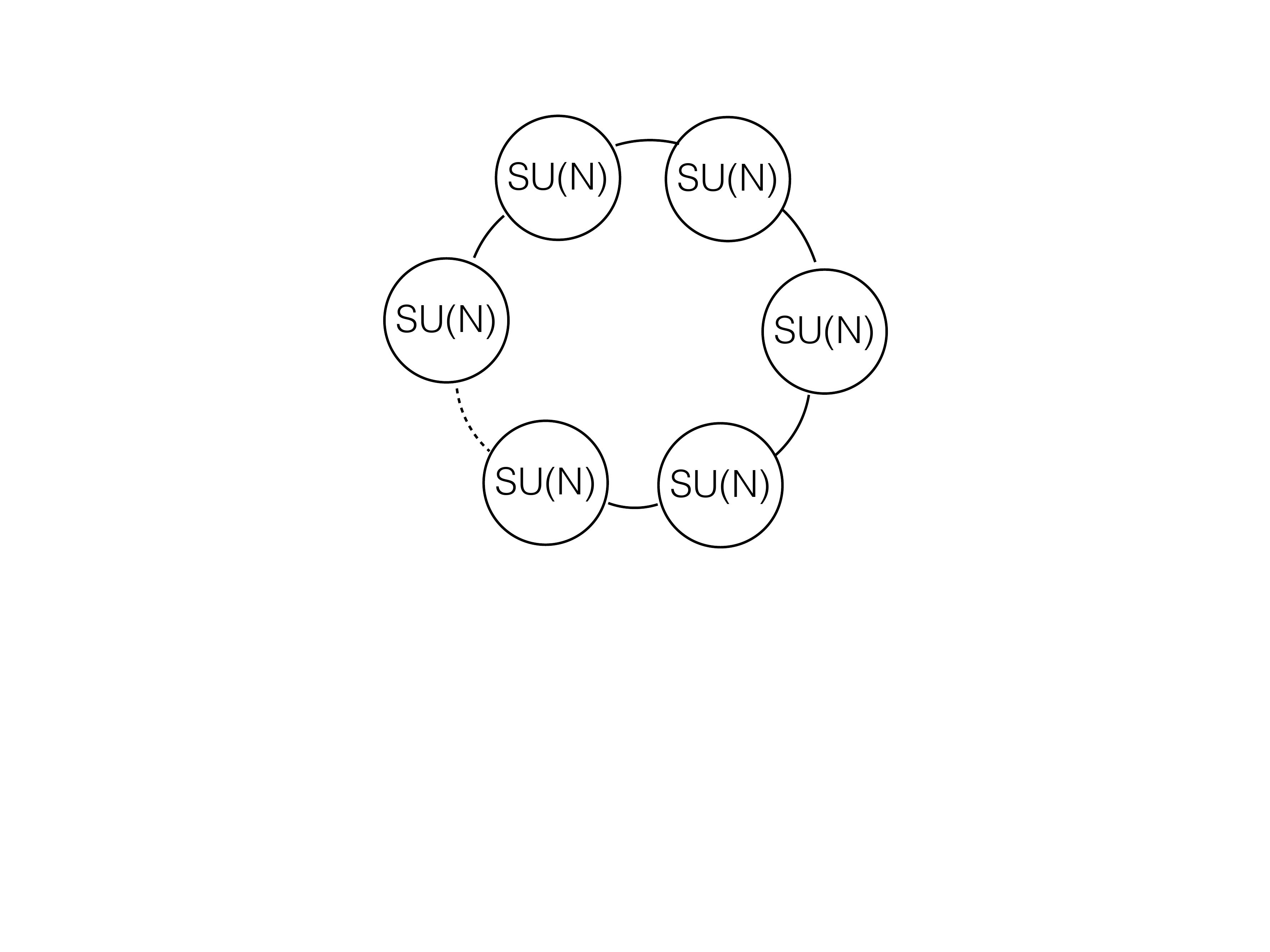}
\vspace{-3cm}
\caption{Circular quiver for $SU(N)^k$}
\label{circular}
\end{figure}

It was also shown in \cite{ArkaniHamed:2001ca} that a sixth extra dimension arises at distances large compared to $a$ as a result of the S-duality symmetry of the 4d $\mathcal{N}=4$ theory: $g_4\leftrightarrow 1/g_4$, which, using  eq.(\ref{gSYM}), implies that $G\leftrightarrow N/G$. Applying this to the spectrum of massive gauge bosons in the latticised 5d theory (also approximated for large $k$):
\begin{equation}
m_n^2=4 G^2 \Phi^2 \sin^2{\frac{\pi n}{k}}\sim \frac{n^2}{R_5^2}\, ,
\end{equation}
new states arise with masses
\begin{equation}
m_n^2=4 \frac{N^2 \Phi^2}{G^2}\sin^2{\frac{n\pi}{k}}\sim  \frac{n^2}{R_6^2}
\end{equation}
for $2\pi R_6=G/\Phi=g_5^2$. Note that under S-duality $R_5\leftrightarrow R_6$. For finite $a<< R_5, R_6$ the quiver in Figure \ref{circular} provides then a discretisation of a 6d theory, which turns out to be the $(2,0)$ CFT living in $N$ M5-branes \cite{ArkaniHamed:2001ie}.

After this brief summary of the ideas of 
\cite{ArkaniHamed:2001ca,ArkaniHamed:2001ie},
we will provide an explicit string theoretical realisation of them.
Indeed,  the $\mathcal{N}=2$  quiver  in  \cite{ArkaniHamed:2001ie}
is the circular quiver that we proposed as dual to our Abelian geometry in eq.(\ref{ads5xs5td}),
 for $\psi$ in a $[0,k\pi]$ interval.
The 4d field theories with gauge coupling $G$ live in the worldvolumes of
D4-branes stretched between NS5-branes, which are therefore wrapped on
$\psi$-intervals of length $\pi$. The effective lattice spacing is 
$a=\pi$.  Using eq. (\ref{couplingd4}), we find for $G$, the gauge
coupling of each of the $SU(N)$ nodes,
\begin{equation}
\frac{1}{G^2}=\frac{1}{2\pi^2}\int_{k\pi}^{(k+1)\pi} d\psi=\frac{1}{2\pi}\, .
\end{equation}
This is the coupling of  each node in the microscopic theory.
At low energies compared to the inverse lattice spacing, the theory should
reduce to a 4d CFT with coupling
$g_4^2=G^2/k$. This corresponds to the gauge coupling of the 4d theory
living on the world-volume
of D4-branes wrapped on the whole circular dimension of length $ka=k\pi$.
Indeed,  at low energies after the Higgs mechanism occurred, the gauge
group is a diagonal combination of all of the microscopic
gauge groups. We can then calculate this coupling from eq.
(\ref{couplingd4}) for $n=k$ to be,
\begin{equation}
\label{g4abelian}
\frac{1}{g_4^2}=\frac{1}{2\pi^2}\int_0^{k\pi}d\psi=\frac{k}{2\pi}=\frac{k}{G^2},
\end{equation}
in agreement with eq.(\ref{gSYM}).
The five dimensional CFT at intermediate energies, appears when
considering a mode expansion of the  fundamental fields (vector and hypers
in the quiver  of Figure
\ref{circular}), that will have periodic boundary conditions on the
$\psi$-coordinate, a calculation that was carefully explained in
\cite{Lambert:2012qy}. This  also suggests that in the Abelian case, the
$\psi$-direction represents the 'theory space' of the quiver in Figure
\ref{circular}.

Let us now try to find a similar interpretation for the non-Abelian quiver
in Figure \ref{linear2}. The situation is a bit more
complicated, because our proposed quiver is not periodically identified.
But a  similar reasoning suggests that a higher dimensional theory
emerges. Indeed, each of the gauge groups in the linear quiver in Figure
\ref{linear2} has a different gauge coupling that can be computed from the
fluctuations of D4-branes wrapped on $r\in[k\pi,(k+1)\pi]$, as in 
eq.(\ref{nonabcou}):
\begin{equation}
\frac{1}{g_{D4,NA}^2}=\frac{1}{G_k^2}=\frac{1}{2\pi^2}\int_{k\pi}^{(k+1)\pi}
r dr=\frac{1}{4}(2k+1)\approx\frac{k}{2}.
\end{equation}
This is  associated with the coupling of each of the nodes of the
microscopic CFT.
If we now let $r$ vary in the whole $[0,k\pi]$ interval we find that, in
the same spirit as above, after Higgsing,
the theory would reduce, at low energies, to a 4d CFT. The gauge group is
a diagonal combination of the microscopic ones, and the  gauge coupling of this IR CFT
is,
\begin{equation}
\frac{1}{g_4^2}=\frac{1}{2\pi^2}\int_0^{k\pi}r dr=\frac{k^2}{4}\approx
\frac{k}{2G_{k}^2}
\end{equation}


Again, this follows the same logic as the Abelian quiver, in this case the
$r$-direction should be the theory space direction. Similar to the Abelian case, one might also argue that the non-Abelian quiver deconstructs two extra directions, hence relating them to the theory on five-branes.

Related ideas that are likely to be useful in a more careful
treatment of the material in this section, have been discussed in
\cite{Sfetsos:2001qb}.

\section{Conclusions and future directions}\label{concl}
Let us briefly summarise, draft some conclusions and speculate about further directions to develop.

We have presented a proposal for the dual CFT to the background obtained by  
non-Abelian T-duality on $AdS_5\times S^5$. Our CFT preserves ${\cal N}=2$ SUSY and fits in  the
description of \cite{Lin:2004nb},\cite{Gaiotto:2009gz}, as a 'long quiver' CFT. We have 
put forward an interesting relation between the Abelian and non-Abelian backgrounds that also works on the field theory side, as a mapping between observables.
 The intuition is that a non-Abelian observable
is related to its Abelian counterpart by a 'discrete differentiation'  along the T-dual coordinate (called $\psi$ or $r$ in this work). 
This connection is valid to leading order for large values of $r$. It suggests that non-Abelian T-duality can be 'continued' at large distances by its Abelian counterpart.
The solution in \cite{Maldacena:2000mw} provides an explicit expression for this 'continuation'.
We have also shown that the Gaiotto-Maldacena formalism can be used on our backgrounds to provide this continuation and smooth them out.

We applied this logic to important observables in the 4d CFT: central charge, Entanglement Entropy and 't Hooft couplings. They follow the connection above mentioned
between the Abelian and non-Abelian results. We found precise four dimensional ${\cal N}=2$ SUSY preserving quivers
matching the values of these observables. 

Reversing the logic, the field theory we proposed suggests a picture in which Abelian T-duality could be 'completed' 
for small values of the dual coordinate by its non-Abelian counterpart. This is in line with what we discussed in Section \ref{Gaiotto-Maldacena}, where the solution 
in \cite{Maldacena:2000mw} was shown to interpolate between the non-Abelian (small $\eta$) and the Abelian (large $\eta$) backgrounds for $\sigma=0$.
This can be seen as a realisation of the idea in \cite{Polychronakos:2010hd}, that non-Abelian T-duality provides a zoom-in on some part of a globally well-defined background,
in this case the one in reference \cite{Maldacena:2000mw}. We presented precise formulas showing this.

In the purely Abelian set-up we showed that the (Hopf) Abelian T-dual of $AdS_5\times S^5$ fits in the classification in \cite{Gaiotto:2009gz} of $\mathcal{N}=2$ geometries, thus providing a connection between $\mathbb{Z}_n$ orbifolds (including the trivial case $n=1$) of $\mathcal{N}=4$ SYM and field theories living in M5-branes. This solution provides an explicit example in which the central charge scales with a $N^2$ power even if associated to M5-branes. $N^\alpha$ scalings with $\alpha\neq 3$ in candidate dual field theories living in M5-branes have been reported before in the non-Abelian T-duality literature (see for instance \cite{Lozano:2015cra}). Here we have shown that it is a common feature also present in more standard Abelian T-dual backgrounds, that remains to be fully understood. We have completed our analysis of the Abelian T-dual solution with a precise realisation of the theory space in 
\cite{ArkaniHamed:2001ca}, from where the 6d $\mathcal{N}=(2,0)$ CFT is deconstructed, in terms of the Type IIA quiver describing the Abelian T-dual of $AdS_5\times S^5/\mathbb{Z}_n$.

In the non-Abelian set up, we have seen that different possible quivers match precisely the 
central charge in eq.(\ref{centralnatdfinal}). It is very likely that these descriptions 
are related through dualities as it happens for some of the quivers proposed in \cite{Gaiotto:2009gz}. This is currently under investigation \cite{LMN}. Similarly, we believe that it would be worth  to further study the deconstruction of a 6d CFT from the non-Abelian T-dual solution, where a six dimensional theory seems to emerge directly at low energies, after Higgsing our proposed quiver. 

An interesting outcome of our studies is the mapping between backgrounds generated by T-duality
and Gaiotto-Maldacena geometries.
The strongly coupled conformal dynamics of these long linear quivers
as well as their deformations and ensuing RG-flow can then be studied
using the dual description that non-Abelian T-duality provides. In cases
of flows to confining field theories or CFTs with less SUSY, modern field
theoretical techniques do not apply and the dual geometry seems to be the
only present tool to tack these problems.

Let us present now, a set of interesting open problems that could be addressed following the developments in this paper.

It would be interesting to find a precise CFT description for the 
backgrounds obtained via non-Abelian T-duality
in other $AdS$ geometries.  The $AdS_5\times T^{1,1}$ case seems the most accessible and promising, but
extensions to $AdS_4\times \Sigma_6$, $AdS_3\times \Sigma_7$, $AdS_6\times \Sigma_4$ cases should also work out nicely. It is expected that the relations between non-Abelian and Abelian T-dual solutions found in this work, as well as our field theory interpretation for the non-Abelian T-dual in terms of a long quiver, will also be applicable in these cases. This would allow to explain in more generality the interplay between non-Abelian T-duality and AdS/CFT. Indeed, the reader familiar with
the papers \cite{Gaiotto:2014lca}, should appreciate the parallelism between their  $AdS_7$ case study and our $AdS_5$ example. It is then clear that it should exist a common formalism in these and many other examples.

It would also be interesting to extend our field theoretical study to geometries that flow to a 'confining background' (or backgrounds representing a flow between CFTs). This would provide
a holographic description of the low-energy confining (or conformal) phase that appears by deforming some of the CFTs above. The 'geometric' side of the work was done in 
some of the papers in \cite{varios1},\cite{varios3}. The present work gives tools to complement that study with a more precise QFT description. 
This will also suggest how to extend the CFTs to the confining phase, this is something quite hard to achieve with present field theoretical techniques.

Furthermore,
it would be nice to study the spectra of the fluctuations of our
probe D4 branes (mesons).
Particularly interesting would be to check if the connection with
deconstruction we have explained, implies that the strongly coupled spectrum gets a KK-like
behaviour or more interestingly even a
$\sin^2(\frac{k\pi}{N})$, as predicted by the weak-coupling analysis done
in deconstruction. 

We hope to report on these and other subtle problems in the future.

\subsection*{Acknowledgements} 
Discussions with various colleagues helped us to improve the contents and presentation of this paper.
We wish to thank: Thiago Araujo, Carlos Hoyos, Georgios Itsios, Niall Macpherson, Horatiu Nastase,  Jes\'us Montero, Leopoldo Pando Zayas, Kostas Sfetsos, Daniel Thompson, Catherine Whiting, Salom\'on Zacarias and very especially Diego Rodriguez-Gomez.
Y.L. is partially supported by the FC-15-GRUPIN-14-108 Project of the Principality of Asturias.  Y.L. and C.N. are partially supported by the EU-COST Action MP1210.
C. Nunez thanks the members of the Physics Department of the University of Oviedo for the  hospitality and nice research environment provided.
C. Nunez is Wolfson Fellow of the Royal Society.

\end{document}